\documentclass[prodmode]{acmsmall} 

\usepackage[ruled]{algorithm2e}
\renewcommand{\algorithmcfname}{ALGORITHM}
\SetAlFnt{\small}
\SetAlCapFnt{\small}
\SetAlCapNameFnt{\small}
\SetAlCapHSkip{0pt}
\IncMargin{-\parindent}

\usepackage{amsmath,amssymb,amsfonts}
\usepackage{bm}
\usepackage{booktabs}
\usepackage{cite}
\usepackage{float}
\usepackage{multirow}
\usepackage{layouts}
\usepackage{pifont}
\usepackage{subfig}
\usepackage{url}
\usepackage[usenames,dvipsnames]{xcolor}
\usepackage{wrapfig}
\usepackage{color}

\acmVolume{1}
\acmNumber{1}
\acmArticle{01}
\acmYear{2014}
\acmMonth{10}

\def\subfigure{\subfloat}

\newcommand{\ch}[1]{{\color{black}{#1}}}
\newcommand{\cut}[1]{{\color{black}{}}}
\newcommand{\ju}[1]{{\color{black}{#1}}}
\begin{document}
\markboth{F. Figueiredo et al.}{On the Dynamics of Social Media Popularity:  A YouTube Case Study}

\title{On the Dynamics of Social Media Popularity: A YouTube Case Study}
\author{FLAVIO FIGUEIREDO,  JUSSARA M. ALMEIDA,  
MARCOS ANDR\'{E} GON\c{C}ALVES and \\
FABR\'{I}CIO BENEVENUTO
\affil{Universidade Federal de Minas Gerais, Brazil}}

\begin{abstract}
Understanding the factors that impact the popularity dynamics of social media   can drive the design of effective information services, besides providing valuable insights to content generators and online advertisers. Taking YouTube as case study, we analyze how video popularity evolves since upload, extracting popularity trends that characterize groups of videos.  We also analyze the referrers that lead users to videos, correlating them, features of the video and early popularity measures with the popularity trend and total observed popularity the video will experience. Our findings provide fundamental knowledge about popularity dynamics and its implications for services such as advertising and search.
\end{abstract}




\acmformat{Flavio Figueiredo, Jussara M. Almeida, Marcos Andr\'{e} Gon\c{c}alves, Fabricio Benevenuto, 
    2014. On the Dynamics of Social Media Popularity: A YouTube Case Study}

\begin{bottomstuff}
This research is partially funded by the Brazilian National  Institute of Science and Technology for  Web Research  (MCT/CNPq/INCT Web Grant Number 573871/2008-6), and by the authors' individual grants from Google, CNPq, CAPES and Fapemig.
We also thank Caetano Traina, Renato Assun\c{c}\~{a}o, Virgilio Almeida, Elizeu Santos-Neto, Matei Ripeanu, and the anonymous reviewers for discussions on drafts of this work. 
\end{bottomstuff}

\maketitle

\section{Introduction} \label{sec:rg1-intro}

User generated content (UGC) has emerged as the predominant form of online information sharing nowadays. The unprecedented amount of information being produced is one of the driving forces behind the success of the social media phenomenon~\cite{Kaplan2010,Cormode2008}. 
This phenomenon is a shift from the traditional media where, instead of content being produced mostly by a few selected individuals, anyone, in theory, can  produce and share content online. 
However, the ``information overload'' that accompanies the huge amount of social media being produced has its drawbacks. For example, it is ever-so-difficult to find and filter relevant content to oneself.  Nevertheless, some pieces of content (or {\it objects}) succeed in attracting the attention of millions of users, while most remain obscure. This leads to the heavy tailed characteristic of content popularity~\cite{Sinha2007,Clauset2009}, where a few objects  become very popular while most of them attract only a handful of views. 
  {\it What  makes one particular object become hugely popular while the majority receive very little attention?} {\it Which factors affect how the popularity of an object will evolve over time?} These are  major questions in the social media context that drive our present work. 

A plethora of different factors may impact social media popularity, including
the object's content itself, the social context
in which it is inserted (e.g., characteristics of the object's creator and her social neighborhood
or influence zone),   mechanisms used
to access the content (e.g., searching, recommendation),
and specific characteristics of the application that may
promote   the visibility of some objects over the others. Some of these factors  contribute to the rich-gets-richer phenomenon~\cite{Easley2010}, which can partially explain the heavy-tailed nature of content popularity.  Others,  such as links to the object from a popular blog and events in the real world, are   external to the application and still may impact the object's future popularity.  

\ch{
Given the importance of social media on society nowadays, understanding 
the extent to which  these factors impact the popularity of social media and how  popularity evolves over time provides valuable insights for content generators, online advertisers and Internet service providers (ISPs), amongst others~\cite{Conover2013,Moat2013,Preis2013,Mestyan2013,Vakali2012}.
In this work, we aim at investigating how different factors impact popularity dynamics of social media, focusing  on YouTube  as case study. YouTube is currently the most popular video sharing application, with  over 100 hours of video   shared  per minute\footnote{\url{http://www.youtube.com/yt/press/statistics.html}}, and a total estimated number of shared videos that had surpassed 4 billion in early 2012\footnote{\url{http://www.reuters.com/article/2012/01/23/us-google-youtube-idUSTRE80M0TS20120123}}.
It  is a rich application that embeds several mechanisms, such as search, list of related videos, and top lists, that may affect how a video is disseminated, thus impacting its popularity.
}



Thus, we here aim at performing a deep study of the evolution of popularity of user generated videos on YouTube. Towards our goal, we collected a public set of statistics available in the system that provides for each video: (a)  its popularity  as a function of time, and (b) a set of referrers, i.e., links used by users to access the video, along with the number of views for which each referrer is responsible. Given the great diversity of content on YouTube,
 our characterization is done on  three different datasets, namely, popular videos that appear on the world-wide top lists maintained by YouTube; videos that were removed from the system due to copyright violation; and, a dataset of videos sampled according to a random procedure (i.e., random queries). 
Focusing on number of views as popularity metric, our study addresses five questions: \\

\vspace{-.8em}
\noindent {\it Q1 - How early do videos reach the majority of observed views?} we intend to assess how fast a video achieves most of its observed popularity. This is key to determine the  time period during which different information services can benefit more from a video. For example, ad placement services will be more effective if ads are posted on videos {\it  before} most of their views are consumed. Moreover, search engines may misleadingly use observed popularity to favor some videos in their rankings, even when videos are no longer attractive.
Our results show that some videos, such as top  and copyright protected videos, achieve most of their views very early on, whereas videos selected  from random queries tend to take longer to attract most of its observed views. \\

\vspace{-.8em}
\noindent {\it Q2 - Is popularity concentrated in bursts?}
    We want to know  whether video popularity  is concentrated on a few days or weeks. This question complements Q1, offering valuable insights into how quickly the interest in the video raises and vanishes.  Moreover,  knowing the peak potential of a video (based on the most popular day/week) is valuable  for services like advertisement campaigns. We find  that  top  and copyright protected videos tend to experience popularity bursts, with a large fraction of their total observed views concentrated on  single week or a even single day, whereas the popularity of videos selected from random queries tends to be less concentrated.  \\

\vspace{-.8em}
\noindent {\it Q3 - Are there governing trends that characterize common groups of video popularity evolution?} We here aim at bridging  Q1 and Q2 by extracting the popularity trends of common groups of videos. To that end, we make use of a time series clustering algorithm \cite{Yang2011}  to infer the popularity trends. Focusing on videos from top lists and selected from random queries, we find that the same four types of popularity trends are observed in both datasets. One trend consists of videos that tend to remain attractive over time with an always increasing popularity. The other trends account for videos that tend to peak in popularity for a short while, with three different popularity decay characteristics after the peak. \cut{We note that, although the peak-based trends have been previously identified using curve fitting models~\cite{Crane2008}, videos that remain attractive over time have not yet been studied. } \\


\vspace{-.8em}
\noindent {\it Q4 - Which incoming links (or referrers) are more important for video popularity, and how early do they occur?} The previous questions   focus on understanding popularity based only on the popularity time series. Here, we want to know how users reach these videos. There are multiple forms through which users can reach a particular piece of content and, thus, there are multiple driving forces that may impact the popularity of a video. Identifying such forces is crucial for designing more cost-effective content dissemination strategies.  For instance, should a content creator invest time on perfecting the keywords describing their videos (for better search rankings) or focus on campaigning videos in online social networks?  Our results show that internal YouTube mechanisms, such as search engines and related videos, are the most important mechanisms that drive users to content, implying that YouTube itself handles a great power to drive video popularity through its internal mechanisms. \\



\vspace{-.8em}
\noindent {\it Q5 - What are the associations between features related to the video, to its early popularity measures and referrers with the popularity trend (or total observed popularity) of the video?}
We aim at  measuring the associations between features related to the video  (e.g.,  category, upload date, age), early points in the popularity time series and  referrers with the identified popularity trends (Q3) and popularity measures. 
We show that videos that follow the same trend tend to also have similar content (based on video category) and
  referrers. For example, music videos tend to remain popular over time and
are generally found through search engines, while videos related to news tend to
have a small but significant attention period and are found through more diverse
sources (e.g., external websites and viral propagation). Moreover, different features are more correlated
with popularity trends and measures at different moments of the video's lifespan,
motivating the use of some of them to build  popularity prediction models.




This work is a follow up on our previous study of the popularity dynamics of YouTube videos~\cite{Figueiredo2011},  which tackled only the first four questions. We here extend it by  introducing Q5,  completely revisiting  how Q3 is addressed, and extending our analyses related to  Q1, Q2 and Q4 to provide thoughtful discussions about the practical implications of our findings for various services as well as   content producers.

The rest of this paper is organized as follows. The next section presents a discussion on related work, while Section~\ref{sec:data} discusses our data collection methodology.
Section~\ref{sec:rg1-growth} presents our characterization of YouTube popularity curves (Q1 and Q2), while Section~\ref{sec:rg1-ksc}  shows the different popularity trends of
YouTube videos (Q3). Next, Section~\ref{sec:rg1-ref}  characterizes the relative importance of different referrers (Q4), whereas Section~\ref{sec:rg1-corr}  discusses the
correlations between various features and popularity  (Q5). Finally, Section~\ref{sec:rg1-conc} concludes the paper discussing the implications of our results.

\section{Related Work}

In this section, we start  by discussing  studies of user generated content (UGC) popularity that focused mainly on static views of popularity (Section~\ref{sec:related-ugcpop}). We then discuss previous efforts to analyze the temporal evolution of UGC popularity (Section~\ref{sec:related-ugcpopevol}).

\subsection{Static Views of Popularity of UGC} \label{sec:related-ugcpop}


In one of the first studies of YouTube video popularity, Cha \textit{et al.}~\cite{Cha2009}  analyzed popularity distribution, popularity evolution and content characteristics of YouTube and of a popular Korean video sharing service, investigating mechanisms to improve video distribution, such as caching and Peer-to-Peer (P2P) content distribution networks (CDNs). Chatzopoulou {\it et al.}~\cite{Chatzopoulou} analyzed  the correlations between  the popularity of YouTube videos, measured in number of views, and other metrics such as numbers of comments and favorites, finding moderate to strong correlations for older videos  and weaker correlations for younger ones, which implies  that their long term popularity dynamics was still unstable.


More recently, Wattenhofer {\it et al.}~\cite{Wattenhofer2012} analyzed the correlations between the popularity of YouTube videos and properties of various online social networks (OSN) created among users of the system. In particular, they found that  characteristics of  the OSN built from links between YouTube users who comment each others' videos
are more correlated to the popularity of a user's video than to the characteristics of the subscription graph (though such correlation  is strong). This result implies that active community collaboration may have a high  impact on the views a user receives through her videos. Similarly,  Susarla {\it et al.}~\cite{Susarla2011} also showed that subscriber links play an important role on the early popularity of videos.  Borghol {\it et al.}~\cite{Borghol2012}  analyzed the correlations between popularity of YouTube videos and content factors, determined by groups of duplicate videos (or clones). Specifically, the authors correlated observed popularity with current popularity and clone groups, using  a linear regression model, and showed that the introduction of new binary explanatory variables capturing the clone groups improved the regression quality,
which implies  that popularity is {\it related} to content. Similarly, Lakkaraju {\it et al.}~\cite{Lakkaraju2013} showed that  the time of day when a Reddit post is added and its title may have a significant impact on its popularity.

\ch{
Flickr images were also the target of many  studies of UGC popularity. For example, Zwol~\cite{Zwol2007} characterized the distributions of total popularity and  popularity decay  over time of images as heavy-tailed. Other studies focused on folksonomies and tags~\cite{Golder2006,Lerman2006,Marlow2006}, which are also examples of UGC. Marlow {\it et al.}~\cite{Marlow2006} found heavy-tailed  distributions of  tag popularity, where the popularity of a tag was estimated by   number of images it annotates and  number of users who used the tag in their libraries. This result has also been observed when tags are used to annotate other kinds of media, such as videos or text data~\cite{Figueiredo2012}. More recently,  Khosla {\it et al.}~\cite{Khosla2014} compared  the use of  image and social features for  predicting the final popularity values of images. Their results are complementary to our results on Q5. However, the authors do not characterize the long-term popularity trends, as we do. 
}


In common, these studies provide important insights into content popularity in various UGC applications. However, most of them focused on either a static snapshot or  at most a few snapshots. Thus, they did not analyze the long-term popularity growth.

\subsection{Popularity Evolution of UGC} \label{sec:related-ugcpopevol}

The popularity evolution of online content has been the target of more recent studies.  Focusing on YouTube videos, Borghol \textit{et al.} showed how weekly based views can be used to model video popularity, and designed a model to determine the number of videos that may exceed some popularity thresholds~\cite{Borghol2011}. This work was recently revisited by Islam {\it et al.}~\cite{Islam2013}, who showed that the weekly based modeling of popularity  is still valid even years after video upload, but the synthetic model proposed for predicting the distribution of popularity of {\it  a group of videos}  is not. Zhou {\it et al.}~\cite{Zhou}  showed the importance of links to related videos to video popularity. The importance of referrers and  content features
 was also briefly discussed in \cite{Borghol2012}, although the authors were more focused on understanding the impact of cloned content on popularity. We  complement these prior efforts by providing a more comprehensive study of popularity trends and their correlations with  various features of the video, its referrers and early popularity measures.

\ch{
Broxton {\it et al.}~\cite{Broxton2011} analyzed  patterns of viral videos, defined as videos that receive a large fraction of views from OSN applications. The authors developed a method to rank different sources of traffic to videos according to their potential in attracting more views. Brodersen {\it et al}.~\cite{Brodersen2012} made use of the same model to  show that most viral videos, after an initial burst in propagation over a diverse set of geographical regions,  tend to fall back to their region of upload. Previous work also focused on geographical propagation of Twitter data~\cite{Kamath2013,Ferrara2013}. Among other things, authors find that \ju{some cities are trend-setters} (sources of popular memes), while others are trend-consumers (sinks).
}


 Cha {\it et al.}~\cite{Cha2012} analyzed the propagation of  pictures through Flickr's internal social network, finding  that the
popularity of the most popular pictures,  measured in number of favorite marks,   exhibits close to linear growth. They also discussed the importance of social links to
  popularity, showing that about 50\% of favorite marks come from social cascades.  In a different direction, Ratkiewicz~{\it et al.}~\cite{Ratkiewicz2010} analyzed how external events, captured by Google
Trends and local browsing (i.e., university traffic), affect the popularity of Wikipedia articles. 

\ch{
 There have also been  efforts to uncover common popularity temporal {\it patterns} or {\it trends}. Crane and Sornette proposed epidemic models to explain a burst in video popularity
in terms of endogenous user interactions and external events~\cite{Crane2008}, whereas Yang and Leskovec~\cite{Yang2011} proposed a time series clustering algorithm to identify
popularity trends. A  unifying analytical framework of the  trends extracted by those studies was proposed in \cite{Matsubara2012}.  We here employ the algorithm proposed in \cite{Yang2011}  to identify  popularity trends in our datasets. Although some of the trends we identify in our datasets are very similar to those reported  in \cite{Yang2011,Crane2008,Figueiredo2011,Lehmann2012},  one of them, which corresponds to videos that remain  popular over time, has not been detected by any previous study.
}

\ch{
Although the aforementioned efforts provide some insights into the evolution of content popularity, there is still little knowledge about which UGC features (e.g., video, referrer, and popularity features) and system mechanisms (e.g., search) contribute the most to popularity growth. Thus, our analyses, performed separately for videos with different characteristics, greatly build on prior efforts, shedding more light into the complex task of understanding UGC popularity.  This paper complements the studies of Borghol {\it et al.}~\cite{Borghol2012,Borghol2011}, who also studied the popularity dynamics of YouTube videos, but did not characterize different popularity trends nor how various features correlate with these trends, as we do. Similarly, our results complement prior efforts focused on geographical views of popularity~\cite{Brodersen2012,Ferrara2013,Kamath2013}, and on extracting important referrers~\cite{Broxton2011}. Many recent studies target the design of popularity prediction models~\cite{Ahmed2013,Pinto2013,Radinsky2012,Li2013,Weng2013,Jiang2014}, but they mostly exploit only popularity time series and/or OSN propagation. One exception is the work of Jiang {\it et al.}~\cite{Jiang2014}, which exploits  content and social features to predict  the day a video is going to peak. However, the authors do not provide a detailed analysis of the importance of each feature to popularity, as we  do. Our results indicate that the aforementioned models could benefit from using various features as input.}


\section{Data Collection} \label{sec:data}

As our case study, we analyze the following datasets, which are publicly available\footnote{\url{http://vod.dcc.ufmg.br/traces/youtime/}}:

\noindent \textbf{Top}: 27,212 videos from  top lists maintained by YouTube (e.g., most viewed videos, most commented videos).

\noindent  \textbf{YouTomb}: 120,862 videos with copyright protected content, identified by the the  MIT  YouTomb project\footnote{\url{http://youtomb.mit.edu/}}. This is the first effort to characterize copyright protected videos.

\ch{
\noindent \textbf{Random topics}:  24,482 videos collected as result of random queries submitted to YouTube’s search API. To build such queries, we first randomly selected, according to a uniform distribution, an entity from the Yago semantic database~\cite{Kasneci2009}. Yago entities cover topics such as popular movies (e.g., Blade Runner) to common items (e.g., Chair). The (textual) name of the entity was then submitted as a query to the YouTube’s search API, and we selected the most relevant video in the result list. We queried for 30,000 entities and discarded queries with empty results\footnote{We do {\it not} claim our dataset is a random sample of YouTube videos. Nevertheless, for the sake of simplicity, we use the term Random videos to refer to videos from this dataset.}.
}


\begin{figure}[t]
 \centering
 \includegraphics[width=.7\linewidth]{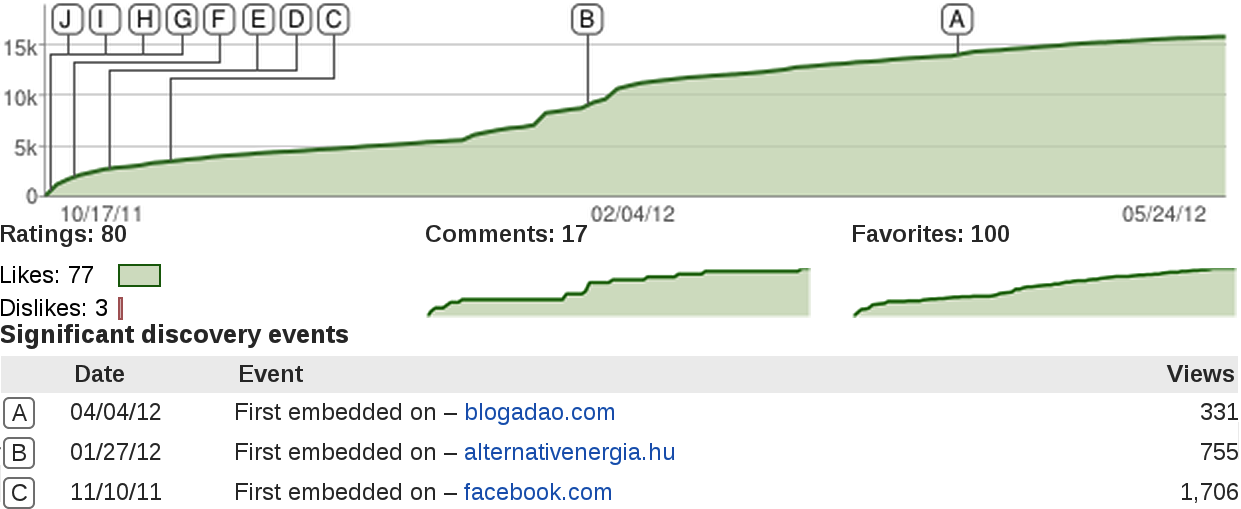}
 \caption{YouTube's insight data example (some referrers were trimmed)}
 \label{fig:ev12}
\end{figure}

\ch{
For each video, we collected YouTube's insight data associated with it, which is publicly available on the video's home page. This insight data consists of various features of the video, including  three time series of how the numbers of views,  comments and favorite markings of the video evolved over time, since the video was uploaded.  It also includes a set of referrers that  led users to the video. The time series are daily for videos with less than 100 days of age, while 100 evenly distributed points are provided   for videos with  more than 100 days of age \cite{Figueiredo2011}. Other features, such as the video category and upload date, were also scrapped from the HTML page of each video. In Figure~\ref{fig:ev12} we show an example of YouTube's insight data that was available up to 2013 (before a change in YouTube's user interface). Currently, the referrer (discovery events) information, comments and favorites time series are no longer provided.
}

We processed our collected datasets  to remove: (1) videos with missing or inconsistent information; and (2)  videos  uploaded on the same day of our crawling. Table~\ref{tab:crawl} provides a summary of each cleaned  dataset,  presenting the total number of videos, average number  of views per video, and average video age\footnote{Changes in our HTML parser lead to datasets of slightly different sizes than our previous study.}. 
Video age, measured in number of days, is defined as the difference between the crawling date (or the removal date, for videos in the YouTomb dataset) and the upload date.  We note that YouTomb videos are on average  older than videos in the Top and Random datasets. Moreover, Top videos are, as expected, more popular, on average, than YouTomb videos, which, in turn, tend  to attract more views than videos in the Random dataset (on average).

We also note that  video ages vary significantly, as shown in Table \ref{tab:ages}. Most videos in the YouTomb and Random datasets are over 1 year old, or have ages between 1 month and 1 year. In contrast, videos in the Top dataset tend to have a bi-modal age range, with most being a few days old or over 1 year. Given such variability, we analyze popularity evolution separately for videos in each age range. However, to avoid hurting presentation with too many graphs, we focus on results computed over all videos in each dataset, pointing out significant differences across age ranges when appropriate.

\begin{table}[ttt!]
\parbox{.48\linewidth}{
\scriptsize
\centering
\caption{Crawled Datasets (after cleanup)}
\begin{tabular}{lllll}
\toprule
Video Datasets & Top & YouTomb & Random \\ 
\midrule
\# of Videos          &  18,422     & 102,888 & 21,935  \\
Avg. \# of of views   &  1,064,264  & 273,696 & 131,473 \\
Avg. video age (days) &  170        & 750     & 526     \\
\bottomrule
\label{tab:crawl}
\end{tabular}
\caption{Distribution of Video Age}
\begin{tabular}{cccc}
\toprule
& Top & YouTomb & Random\\
\cmidrule(l){2-4}
age (days) $\leq$ 7             & 4,303 & 0      & 109      \\
7 $<$ age (days) $\leq$ 30      & 6,543 & 0      & 563      \\
30 $<$ age (days) $\leq$ 365    & 4,627 & 13,379 & 8,159 \\
age (days) $>$ 365              & 2,949 & 89,509 & 13,104 \\
\bottomrule
\end{tabular}
\label{tab:ages}
}\hfill
\parbox{.48\linewidth}{
\scriptsize
\centering
\caption{Summary of Features}
\begin{tabular}{lll} \toprule
Class & Feature Name & Type  \\
\midrule
\multirow{2}{*}{Video}    & Video category           & Categorical   \\
                          & Upload date              & Numerical   \\
			& Video age & Numerical \\
		    & Time window size ($w$)   & Numerical    \\ \midrule
\multirow{2}{*}{Referrer}     & Referrer first date      & Numerical    \\
                          & Referrer \# of views & Numerical   \\ \midrule
\multirow{5}{*}{Popularity} & \# of views              & Numerical   \\
                          & \# of comments           & Numerical     \\
                          & \# of favorites          & Numerical     \\
                          & change rate of views   & Numerical     \\
                          & change rate of comments   & Numerical     \\
                          & change rate of favorites   & Numerical     \\
                          & Peak fraction         & Numerical     \\
\bottomrule
\label{tab:feats}
\end{tabular}}
\end{table}

\ch{
The features we collected, shown in Table \ref{tab:feats},  are grouped into three classes, namely  video, referrer, and popularity features. Video features include category, upload date, age, and the duration of the time window $w$ that represents a single observation in the video's popularity time series (see below). The video category is defined based on the YouTube's list of categories, which includes Autos/Vehicles, Comedy, Education, Entertainment, Gaming, Film/Animation, Howto/Style, Music, News/Politics, Shows, Nonprofit/Activism, People/Blogs, Pets/Animals,  Travel/Events,  Science/Technology, and Sports. The referrer features include the first date and the number of views associated with each referrer category. Referrers are categorized into {\it External, Featured, Search, Internal, Mobile, Social and Viral}. The {\it External} category represents websites (often other OSNs and blogs) that have links to the video. The {\it Featured} category contains referrers that come from advertises about the video in other YouTube pages or featured videos on top lists and on the front page. The {\it  Search} category includes referrers from search engines, which comprise only Google services. {\it Internal} referrers correspond to other YouTube mechanisms, such as  the ``Related Video'' feature.
{\it Mobile} includes all accesses that come from mobile devices. {\it Social} referrers consist of accesses from the page of the video owner  or from users who subscribed to the owner or to some specific topic. Finally, some other referrers are grouped into {\it Viral}. 
The popularity features include the total numbers of views, comments and times the video was marked as favorite, the trend in these measures captured by the corresponding average change rates, and the largest fraction of all observed views that happened in a single time window (peak fraction).  Jointly, these features capture properties of the popularity curve.
}

We note some limitations of the data provided by YouTube. Each popularity curve is registered with at most 100 points, regardless of the video age. Thus, the video's time window $w$ is defined as the video age divided by 100.  In order to be able to estimate video popularity on a daily basis, we performed linear interpolation between the 100 points provided. Moreover, YouTube does not provide information on every referrer that led users to the videos, but rather on  ten {\it important} ones (according to YouTube). In total, the available referrers account for only 36\%, 25\% and 35\% of all the views of  videos in  the Top, YouTomb, and Random datasets, respectively.

\section{Understanding Video Popularity Growth} \label{sec:rg1-growth}

We start by analyzing  the popularity growth patterns of videos in our three datasets,  focusing on two aspects:  (1) the time interval until a video reaches most of its observed popularity, and (2) the bursts of popularity experienced by a video in short periods of times (e.g., days or weeks). 
We use the number of views as popularity metric because previous studies have found  large correlations between total number of comments (or favorites) and total view count~\cite{Chatzopoulou}. Moreover,  we have also found positive correlations, ranging from 0.18 to 0.24, for both pairs of metrics, taken at  each point in time (instead of only for the final snapshot, as previously done).

\subsection{ How early do videos reach most of its observed views (Q1)?} \label{subsec:early}

\begin{figure}[t]
 \centering
 \mbox{\subfloat[Top]{\includegraphics[width=0.3\linewidth]{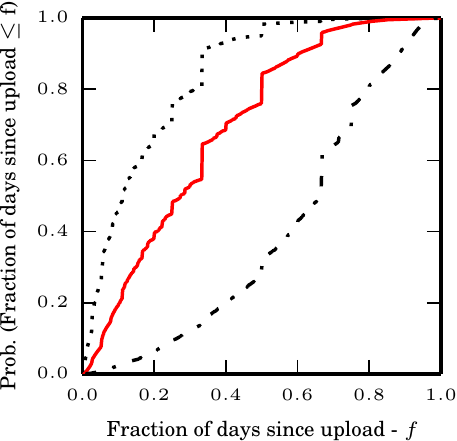}}}
 \mbox{\subfloat[YouTomb]{\includegraphics[width=0.3\linewidth]{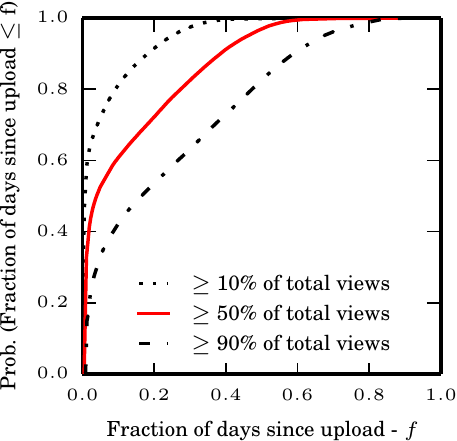}}}
 \mbox{\subfloat[Random]{\includegraphics[width=0.3\linewidth]{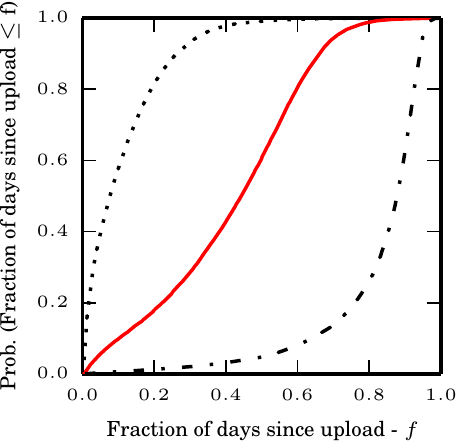}}}
 \caption{Cumulative Distributions of Time Until Video Achieves at Least 10\%, 50\% and 90\% of its Total Observed Popularity  (time normalized by video's lifespan).}
 \label{fig:105090}
\end{figure}

Figure~\ref{fig:105090} shows the cumulative distributions of the amount of time it takes for a video to receive {\it at least 10\%}, {\it at least 50\%} and {\it at least 90\%} of their total (observed) views, measured at the time our data was collected. Time is shown normalized by the age of the video, which is here referred to as the video's {\it lifespan}. \ju{That is, the y-axis shows the  fraction of videos that achieved at least 10\%, 50\%, and 90\% of their total views (considering the total views at the time we crawled the data) in  a period of time that does not exceed the value shown in the x-axis (which is normalized by the total time since the video was uploaded).}


We note that, for half of the videos (y-axis) in the Top, YouTomb and Random datasets, it takes at most 67\%, 17\% and 87\%, respectively, of their total lifespans (x-axis) until they receive at least 90\% of their total views. If we consider at least 50\% of their total views, the fractions are  27\%, 4\%  and 44\%, respectively, following a similar trend (as also found for the mark of 10\% of the views). Conversely, around 34\% of Top videos take at least 20\% of their lifespans to reach at least 10\% of their observed popularity. Similarly, 19\% of    videos in the Random dataset experience a similar dormant period before starting to receive most views.  In contrast, only 8\% of the YouTomb videos take 20\% or more of their lifespans to reach at least 10\%  of their observed popularity.

Thus, comparing the results across datasets, YouTomb videos tend to get most of their views earlier in their lifespans, followed by videos in Top and Random datasets. As videos in the top lists tend to be more popular, the difference between the results for Top and Random datasets are somewhat predictable. Possible reasons as to why YouTomb videos tend to receive most of their views even earlier are: (1) as many of these videos consist of popular TV shows and music trailers, a natural interest in this content closer to when it is uploaded is expected, and (2) being aware that such videos contain copyright protected content, users may seek them quicker after upload,  before the violation is detected and they are removed from YouTube.

We note that since lifespan is a normalized metric, these results may be impacted by the distributions of video ages (Table \ref{tab:ages}). In particular, recall that such distribution is skewed towards older videos in the YouTomb dataset: around 86\% of them have at least 1 year of age. This bias may influence the results. However, we also note that 59\% of the videos in the Random dataset also fall into the same age range. Yet, in comparison with YouTomb, videos in the Random dataset get most of their views later. 

Thus, to reduce any bias caused by age differences, we repeat our analyses separately for videos in each age range. Table \ref{tab:early_ages} shows results for  the time until a video achieves at least 90\% of its views, presenting averages and standard deviations for each age range and dataset. Similar results occur for videos in most age ranges: YouTomb videos reach at least 90\% of their views  much earlier in their lifespans than Top videos,  which are followed by videos in the Random dataset.  The only exception occurs for the youngest videos, for which there is no much difference across the datasets.

\subsection{Is popularity concentrated in bursts (Q2)?} \label{subsec:conce}

\begin{figure}[t]
 \centering
 \mbox{\subfigure[Top (Peak Day)]{\includegraphics[width=0.3\linewidth]{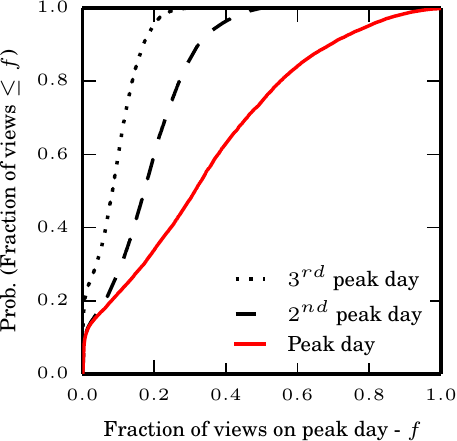}}}
 \mbox{\subfigure[YouTomb (Peak Day)]{\includegraphics[width=0.3\linewidth]{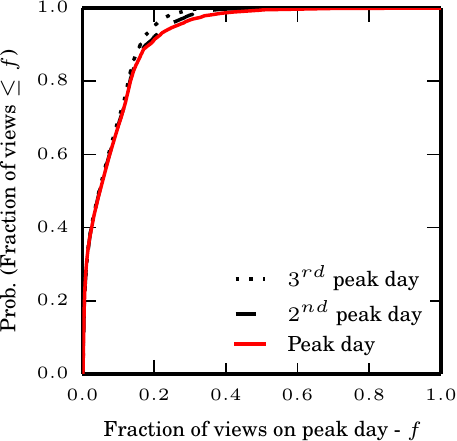}}}
 \mbox{\subfigure[Random (Peak Day)]{\includegraphics[width=0.3\linewidth]{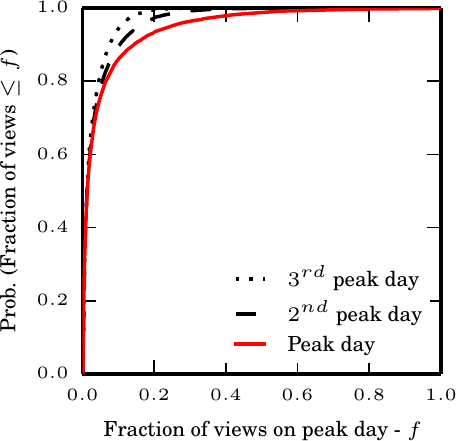}}}
 \mbox{\subfigure[Top (Peak Week)]{\includegraphics[width=0.3\linewidth]{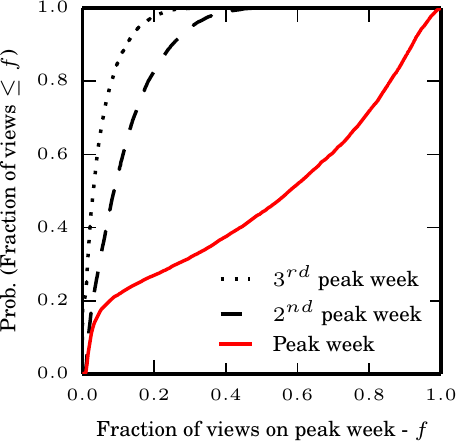}}}
 \mbox{\subfigure[YouTomb (Peak Week)]{\includegraphics[width=0.3\linewidth]{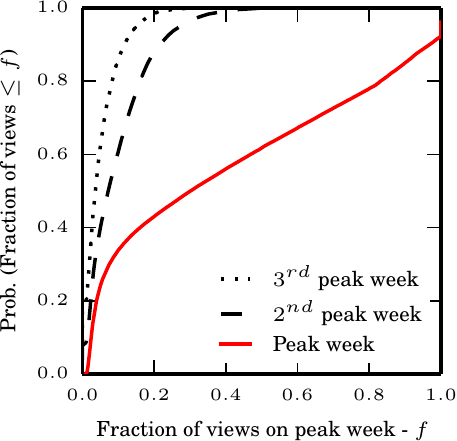}}}
 \mbox{\subfigure[Random (Peak Week)]{\includegraphics[width=0.3\linewidth]{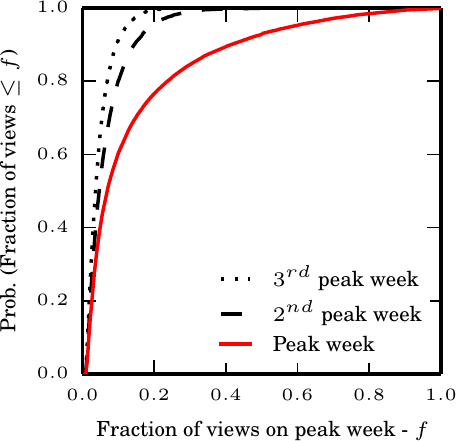}}}
 \caption{Cumulative Distributions of  the Fraction of Total Views on the First, Second and Third Peak Days/Weeks.}
 \label{fig:dpeak}
\end{figure}

We now investigate the popularity bursts experienced by the videos. We first analyze the  distributions of the fraction of views a video receives on its most popular (i.e., peak) day,  shown in  Figure~\ref{fig:dpeak} and summarized in Table \ref{tab:peak} for videos falling in different age ranges. Figure \ref{fig:dpeak} also shows  distributions for the second and third most popular days. \ju{Each curve in a graph of Figure \ref{fig:dpeak} shows the fraction of videos (y-axis) that receive at most $f$\% (shown in x-axis, as a fraction) of its total views on the given peak day. }

\begin{table}[t]
\parbox{.48\linewidth}{
\scriptsize
\centering
\caption{Normalized Time Until at Least 90\% of Total Views, Grouped by Video Age (mean $\mu$, and standard deviation $\sigma$).}
\begin{tabular}{lllllll}
\toprule
 & \multicolumn{2}{c}{Top} & \multicolumn{2}{c}{YouTomb} & \multicolumn{2}{c}{Random} \\
\cmidrule(l){2-7}
 & $\mu$ & $\sigma$ & $\mu$ & $\sigma$ & $\mu$ & $\sigma$ \\
\cmidrule(l){2-7}
age (days) $\leq$ 7            & .64 & .10 &  -  &  -  & .60 & .16 \\
7 $<$ age $\leq$ 30            & .56 & .19 &  -  &  -  & .66 & .21 \\
30 $<$ age $\leq$ 365          & .50 & .27 & .10 & .13 & .80 & .17 \\
age $>$ 365                    & .77 & .23 & .26 & .23 & .85 & .12 \\
\bottomrule
\label{tab:early_ages}
\end{tabular}
}\hfill
\parbox{.48\linewidth}{
\centering
\scriptsize
\caption{Fraction of Views on Peak Day Grouped by Video Age (mean $\mu$, and standard deviation $\sigma$).}
\begin{tabular}{lllllll}
\toprule
 & \multicolumn{2}{c}{Top} & \multicolumn{2}{c}{YouTomb} & \multicolumn{2}{c}{Random} \\
\cmidrule(l){2-7}
        & $\mu$ & $\delta$ & $\mu$ & $\delta$ & $\mu$ & $\delta$ \\
\cmidrule(l){2-7}
age (days) $\leq$ 7       & 64\% & .16 &  -   &  - & 63\% & .20\\
7 $<$ age $\leq$ 30       & 35\% & .15 &  -   &  - & 33\% & .19\\
30 $<$ age $\leq$ 365     & 23\% & .16 & 21\% & .13 & 8\%  & .11\\
age $>$ 365               & 2\%  & .03  & 20\% & .03  & 1\%  & .02\\
\bottomrule
\label{tab:peak}
\end{tabular}
}
\end{table}

Figure~\ref{fig:dpeak}-a) shows that Top videos experience a very distinct peak day: 50\% of them receive between 31\% and  100\% of their views on a single (peak) day. In comparison, the same fraction of videos receive between 17\% and 50\% of their views on the second peak day, and between 8\% and 34\% of their views on their third peak day. Thus, Top videos clearly experience a burst of popularity on a single  day.  This is in sharp contrast with videos in the YouTomb and Random datasets (Figures~\ref{fig:dpeak}-b and \ref{fig:dpeak}c), where the three curves are very close to each other and skewed towards smaller fractions of views.  
While these results might reflect diverse popularity patterns, with more videos in the Random and YouTomb datasets having multiple (smaller) daily peaks,   we note that the interpolation performed over the collected data might introduce distortions in this analysis, particularly given the large fraction of older videos in those two datasets.

To cope with these possible distortions, we also analyze the distributions of the fraction of views on the first, second and third peak weeks. Figures \ref{fig:dpeak}(d-f) show that videos in all  datasets tend to exhibit some burst of popularity on a single week. However, the general trend remains the same as  for daily peaks: the peak week tends to be more significant for Top videos, followed by videos in  the YouTomb and Random datasets.

The same general conclusions, for both weekly and daily popularity peaks, also hold for videos falling in different age ranges,  as illustrated in Table~\ref{tab:peak} for daily peaks.

\subsection{Discussion}

In this section we characterized content popularity growth, focusing on our first two research questions (Q1 and Q2). 
In general, we note that results vary according to the analyzed dataset. While Top and YouTomb videos tend to be more concentrated and receive most views earlier in their lifespans,  videos in the Random dataset  exhibit less clear bursts, particularly at the daily granularity, and tend to take longer to receive most views.  These results contrast and complement previous analyses of YouTube videos, where the authors characterized a sample of videos uploaded on a single day, concluding that they exhibit concentrated popularity growth patterns \cite{Borghol2011}.  By analyzing different  datasets,  composed of videos with different characteristics, our study is able to reveal different aspects of YouTube as a whole. 



These results  might be  useful for a wide range of social media services. For example, they raise the question of whether (and when)  it is beneficial to incorporate popularity estimates into search engine rankings. For  videos that receive most of its views in short time periods (such as videos in the Top and YouTomb datasets), adding this information into the ranking  after the period of interest has already passed might hide other (possibly more relevant) videos (e.g., newly uploaded videos). Another interesting argument is for advertisement services. The notion that popular content may have a higher ad-visibility has been  discussed only  recently~\cite{Carrascosa2013}.  However, focusing on the final observed popularity may be misleading, since posting ads on popular videos does not necessarily promotes a higher amount of {\it future audience}.

\section{Popularity Temporal Dynamics (Q3)} \label{sec:rg1-ksc}

We now characterize the temporal dynamics of popularity of YouTube videos, aiming at identifying governing popularity trends that characterize groups of videos in our datasets. To that end, 
we employ the KSC algorithm~\cite{Yang2011}, which  is a K-Means like clustering algorithm focused on extracting similar trends (or shapes) from time series. KSC is based on a distance metric that captures the similarity between two time series with scale and time shifting invariants. 

 KSC requires all time series to have equal length. Thus, we   focus on videos with more than 100 days, whose popularity time series is defined by 100 evenly distributed observations, that is, the original crawled data with no interpolation\footnote{\ju{The popularity curves of those videos capture longer term popularity dynamics and trends.}}. Each such observation represents the popularity of the video at a time window $w$, whose duration depends on the video age. We also focus  on the Top and Random datasets,  since the non-interpolated data from the YouTomb dataset has all zeros after the removal date, which leads to  time series with various lengths that cannot be handled by KSC. After such filtering, we are left with 4,527 and  19,562 videos in the Top and Random datasets, respectively. These are the videos analyzed in this section (and  in Section \ref{sec:rg1-corr}).

Like K-means, the KSC algorithm requires the target number of clusters $k$ to be given as input. We  use the $\beta_{CV}$ heuristic \cite{Menasce2002} to define the best value of $k$.  The $\beta_{CV}$ is defined as the ratio of the coefficient of variation  (CV) 
of the intracluster distances and the coefficient of variation
of the intercluster distances. The smallest value of $k$ after which the $\beta_{CV}$ remains roughly stable should be selected, as a stable $\beta_{CV}$
implies that new splits affect only marginally the variations of intra and intercluster distances.
The values of $\beta_{CV}$ seem to stabilize for $k$=$4$, for both analyzed datasets.
We confirmed this choice by  plotting the clustering cost, silhouette and Hartigan's index metrics~\cite{Yang2011}, and by visually inspecting the  members of each cluster. The best value of $k$ was 4 according to all these techniques.

\begin{figure}[t]
 \centering
 \includegraphics{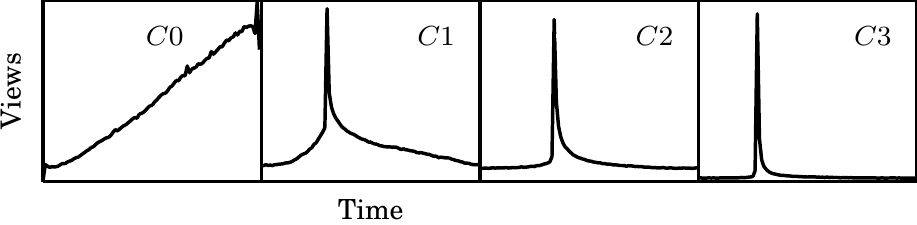}
 \caption{ Popularity Trends (Cluster Centroids) in Both Random and Top Datasets.}
 \label{fig:cent-rand}
\end{figure}

Figure \ref{fig:cent-rand} shows the discovered popularity trends (the centroids of the identified clusters), which govern popularity evolution in our datasets. Each graph shows the number of views as function of time. Note that the same four popularity trends are present in both analyzed datasets. Moreover, Table~\ref{tab:summclus} presents, for each cluster, the number of videos that belong to it as well as the average  number of views, the average change rate in the number of views, and the fraction of views at the peak time window of these videos.   The  average change rate is the average difference between two (non-cumulative) measures taken in successive time windows. Thus, it captures the trend in the number of views of the video: a positive (negative) change rate indicates an increase (decrease) with time, whereas a change rate equal to 0 indicates stability.  Table \ref{tab:summclus}  shows the average change rate computed over the total duration of the video's lifespan.  The peak fraction, also shown in Table \ref{tab:summclus}, is the ratio of the maximum number of views in a time window divided by the total number of views of  the video.

\begin{table}[t]
\scriptsize
\centering
\caption{Summary of Popularity Trends}
\begin{tabular}{l*{3}{c}c|*{4}{c}} \toprule
& \multicolumn{4}{c}{Top} & \multicolumn{4}{c}{Random} \\
\cmidrule(r){2-9}
& $C0$ & $C1$ & $C2$ & $C3$ & $C0$ & $C1$ & $C2$ & $C3$ \\
\cmidrule(r){2-9}
Number of Videos       & 958     & 1,370     & 1,084    & 1,115    & 4,023 & 6,718 & 5,031 & 3,790\\
Avg. Number  of Views   & 711,868  & 6,133,348  & 1,440,469 & 1,279,506 & 305,130 & 108,844 & 64,274 & 127,768\\
Avg.  Change  Rate  & 1112 & 395 & 51 & 67  & 47 & 7 & 4 & 4\\
Avg. Peak Fraction & 0.03    & 0.04     & 0.19 & 0.74 & 0.03 & 0.03 & 0.08 & 0.28\\
\hline
\end{tabular}
\label{tab:summclus}
\end{table}

\ch{
As shown in Figure~\ref{fig:cent-rand}, cluster 0 consists of videos that remain popular over time,  attracting an increasing number of views per time window as time passes, as indicated by the large positive  change rates (Table \ref{tab:summclus}).  This behavior is specially strong in the Top dataset, with an average change rate of 1,112 views per window,  which  corresponds to roughly a week in that dataset. The videos in cluster 0 have also no significant peaks, as the average fractions of views in the peak windows are very small (Table \ref{tab:summclus}).  
The other three clusters are predominantly defined by a single peak in popularity followed by a steady decline. The main difference is the rate of decline, which is much slower in Cluster 1, somewhat faster in Cluster 2, and very sharp in Cluster 3. This difference is more clear if we analyze the peak fractions and the average change rates in Table~\ref{tab:summclus}. }

\ju{
Given the popularity (i.e., scale) invariant nature of the KSC algorithm, it is important to highlight the differences between the clusters in the Top and Random datasets. To that end, we make use of the numbers in Table \ref{tab:summclus}. Although very similar clusters exist in both datasets (determined both by the shape of the centroids and the fraction of videos in each cluster), notice that the change rates in popularity for the videos in the Top dataset are much higher (for every cluster) than the corresponding rates in the Random dataset. For example, videos in Cluster 0 (which remain popular over time) in the Top dataset experience a change in number of views in consecutive time windows of 1,112 views, on average. In contrast, videos in the Random dataset experience a change of only 47 views, on average. 

Also notice how the peak fractions in the Top dataset are higher than those in the Random dataset (in all clusters but Cluster 0). However, the average number of views in Cluster 0 in the Top dataset is the smallest one when compared to the other clusters in the same dataset. For the Random dataset, this is the opposite. This is very interesting, as it indicates that  the most popular videos in the Top dataset are in Clusters 1-3, that is, they experience clear popularity peaks, being more popular in  shorter time windows. However, given the very high change rates experienced by videos in Cluster 0 (in Top), we might speculate that videos in this cluster will become more popular over time, as they capture enough interest to remain receiving visitors over time. We might also speculate that, as time passes and the Top videos in Clusters 1-3 loose their appeal to the audience, the relative distribution of popularity across clusters in the Top dataset will be more similar to that in the Random dataset. This is a conjecture that requires further investigation in the future.
}

It is also important to note that Clusters 1, 2 and 3 were previously uncovered in other YouTube or Twitter datasets \cite{Crane2008,Lehmann2012,Matsubara2012}. Crane and Sornette \cite{Crane2008} explained their occurrences by a combination of endogenous user interactions and external factors. According to them, Cluster 1 consists of videos that experience word-of-mouth popularity growth resulting from epidemic-like propagation through the social network; Cluster 2 includes videos that experience a sudden popularity burst, due to some external event, but continue spreading through the social network afterwards; and Cluster 3 consists of videos that experience a popularity burst for some reason (e.g., spam) but do not spread through the social network. However, these previous studies relied mostly on peak popularity analyses~\cite{Lehmann2012} and fitting power-law decays after the peak~\cite{Matsubara2012,Crane2008}. Instead, we here use an unsupervised learning algorithm that makes our task of discovering popularity trends more general and robust. For example, the thresholds in peak volume that define different trends in these previous studies are not clearly defined. In contrast, such peaks emerge clearly in our clusters (as shown in Table \ref{tab:summclus}). 

Notice however that no previous study that analyzed video popularity time series or other UGC time series has identified a trend similar to Cluster 0, possibly because of the models they adopted, which focus on power-law like behavior \cite{Crane2008,Matsubara2012}  or due to inherent differences in media consumption trends for different media types \cite{Yang2011}. The existence of Cluster 0 can be attributed to three possible reasons. Firstly, there are certain topics that users will continue to revisit over time~\cite{Anderson,Wang2012}, and thus the content will not follow a rise-and-fall pattern (as proposed in \cite{Matsubara2012}). Secondly, the propagation of these topics is much slower~\cite{Wang2012}, being the pattern we see still part of the growth period in interest in that particular topic. Lastly, YouTube's own growth in popularity over time may cause the audience of interest in some videos to increase. Intuitively, a combination of these factors will likely be the case, and only recently researchers have started looking into the implications of each of them~\cite{Anderson,Wang2012}. 

\ju{Finally, we note that other time series clustering techniques could also be employed to extract
 popularity trends from our datasets. For example, one could consider first using  {\it Symbolic Aggregate Approximation}, SAX  \cite{Lin2007}  to represent the time series, and then applying traditional clustering methods (e.g., K-Means). However, SAX  assumes that time series values are normally distributed, which is not true for our data (even after log and z-transformations). We argue that KSC is a suitable choice of clustering algorithm to our  study because  it: (1) requires only  the number of clusters as input, (2) requires no data pre-processing, and (3) has well defined and interpretable centroids, which facilitates  analyzing and drawing useful insights from the results.}

This section bridges our study on Q1 and Q2, and thus have similar implications for social media services. So far we  characterized video popularity focusing on popularity time series only. We have yet to discuss possible reasons behind content popularity and popularity trends. We explore these issues in the next two sections.
Throughout the rest of the paper, we refer to Clusters 0, 1, 2 and 3 as $C0$, $C1$, $C2$, and $C3$, respectively.




\section{Referrer Analysis (Q4)} \label{sec:rg1-ref}

The dynamics of  information propagation through friends in social networks has been studied before \cite{Cha2012}. However, on YouTube,  as on  other social media applications, word-of-mouth is not the only mechanism through which information is disseminated. We here tackle this issue by  investigating  important referrers that lead users to videos (Section \ref{subsec:Q4a}) and their first access since video upload (Section \ref{subsec:Q4b}). These analyses are performed on our three original datasets (Table \ref{tab:crawl}).

\subsection{ Which referrers are more important for video popularity (Q4a)?} \label{subsec:Q4a}

\begin{table}[t]
\centering \scriptsize
\caption{Referrer  Statistics ($n_{view}$: number of views (x $10^9$); $f_{view}$:  fraction of 
views; $f_{time}$: fraction of times a referrer from the category was the first referrer of a  video).}
\begin{tabular}{lp{4cm}p{0.3cm}p{0.3cm}p{0.3cm}p{0.3cm}p{0.3cm}p{0.3cm}p{0.3cm}p{0.3cm}p{0.3cm}p{0.3cm}p{0.3cm}p{0.3cm}}
\toprule
\vspace{-0.3cm}
 &  & \multicolumn{3}{c}{Top} & \multicolumn{3}{c}{YouTomb} & \multicolumn{3}{c}{Random}\\
\multirow{2}{*}{Category} & \multirow{2}{*}{Referrer Type} \\
\cmidrule{3-11}
 &  & $n_{view}$ & $f_{view}$ & $f_{time}$ & $n_{view}$ & $f_{view}$ & $f_{time}$ & $n_{view}$ & $f_{view}$ & 
$f_{time}$ \\
\midrule
\multirow{3}{*}{EXTERNAL} &   First embedded view &
\multirow{3}{*}{0.57} & \multirow{3}{*}{0.11} & \multirow{3}{*}{0.35} & 
\multirow{3}{*}{0.81} & \multirow{3}{*}{0.16} & \multirow{3}{*}{0.41} & 
\multirow{3}{*}{0.07} & \multirow{3}{*}{0.08} & \multirow{3}{*}{0.22} \\
                          &   First embedded on   \\
                          &   First referrer from \\
                          \midrule
\multirow{2}{*}{FEATURED} &   First view from ad &
\multirow{2}{*}{0.72} & \multirow{2}{*}{0.14} & \multirow{2}{*}{0.03} & 
\multirow{2}{*}{0.10} & \multirow{2}{*}{0.02} & \multirow{2}{*}{0.00} & 
\multirow{2}{*}{0.11} & \multirow{2}{*}{0.14} & \multirow{2}{*}{0.00} \\
                          &   First featured video view \\
                          \midrule
\multirow{2}{*}{INTERNAL} &   First referrer from YouTube &
\multirow{2}{*}{1.50} & \multirow{2}{*}{0.29} & \multirow{2}{*}{0.67} & 
\multirow{2}{*}{1.85} & \multirow{2}{*}{0.36} & \multirow{2}{*}{0.65} & 
\multirow{2}{*}{0.14} & \multirow{2}{*}{0.18} & \multirow{2}{*}{0.34} \\
                          &   First referrer from Related Video \\
                          \midrule
\multirow{1}{*}{MOBILE}   &   First view from a mobile device & 
0.26 & 0.05 & 0.51 & 
0.02 & 0.00 & 0.02 & 
0.03 & 0.03 & 0.05 \\
                    	\midrule
\multirow{3}{*}{SEARCH}   &   First referrer from Google &
\multirow{3}{*}{1.05} & \multirow{3}{*}{0.20} & \multirow{3}{*}{0.36} & 
\multirow{3}{*}{1.80} & \multirow{3}{*}{0.35} & \multirow{3}{*}{0.52} & 
\multirow{3}{*}{0.29} & \multirow{3}{*}{0.37} & \multirow{3}{*}{0.41} \\
                          &   First referrer from YT search \\
                          &   First referrer from Google Video   \\
                          \midrule
\multirow{2}{*}{SOCIAL}   &   First referrer from a subscriber & 
\multirow{2}{*}{0.36} & \multirow{2}{*}{0.07} & \multirow{2}{*}{0.35} & 
\multirow{2}{*}{0.01} & \multirow{2}{*}{0.00} & \multirow{2}{*}{0.01} & 
\multirow{2}{*}{0.01} & \multirow{2}{*}{0.00} & \multirow{2}{*}{0.12} \\
                          &   First view on a channel page \\
                          \midrule
\multirow{1}{*}{VIRAL}    &   Other / Viral & 
0.81 & 0.16 & 0.79 &  
0.59 & 0.12 & 0.62 & 
0.16 & 0.20 & 0.55 \\                          
\bottomrule
\end{tabular}
\label{tab:groups}
\end{table}

Recall that the referrers  in our datasets were grouped into seven categories: External, Featured, Search, Internal, Mobile, Social, and Viral. Table~\ref{tab:groups} shows the number ($n_{view}$) and fraction ($f_{view}$) of views for which each category is responsible.   The table shows that  search and internal YouTube mechanisms are key channels through which users reach content on the system, and  we note that YouTube search is responsible for more than 99\% of all Search referrers.   Oliveira \textit{et al.}~\cite{Oliveira2010} posed the hypothesis that search is the main method for reaching content on video sharing websites, verifying it through questionnaires with volunteers.  Whereas our results  confirm their hypothesis for videos in the Random dataset, we find that  YouTube internal features (e.g., ``Related Videos")  play an even more important role to content dissemination for Top videos. For YouTomb videos, both categories are roughly equally important. 
In general, we find that
search is more
important to Random and YouTomb videos, as they are not systematically exposed to users as videos from top lists are. 
We also note the importance of the Viral  category in all  datasets, particularly Random.

\begin{figure}[t]
 \centering
 \mbox{\subfloat[Top]{\includegraphics[width=0.3\linewidth]{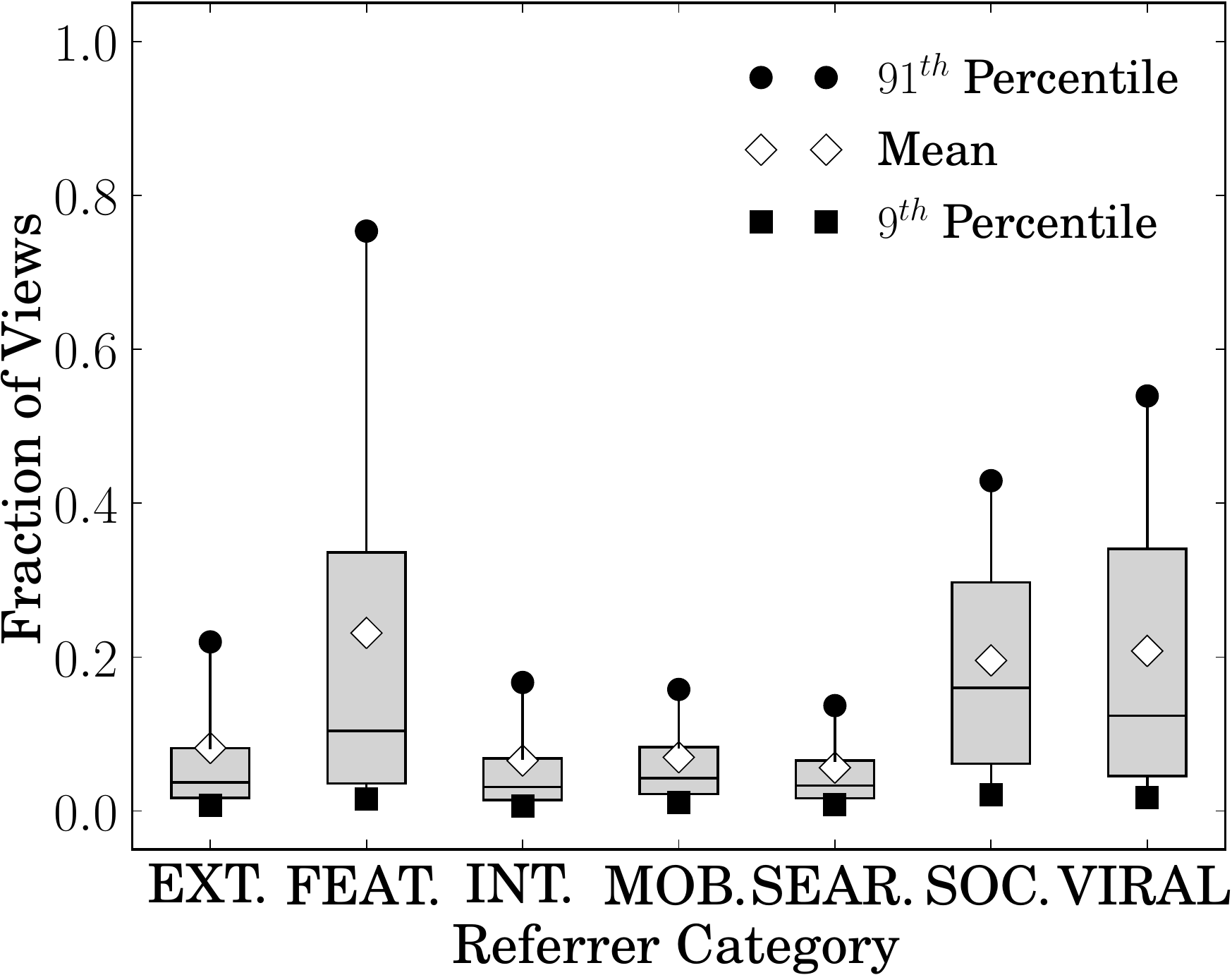}}}
 \mbox{\subfloat[YouTomb]{\includegraphics[width=0.3\linewidth]{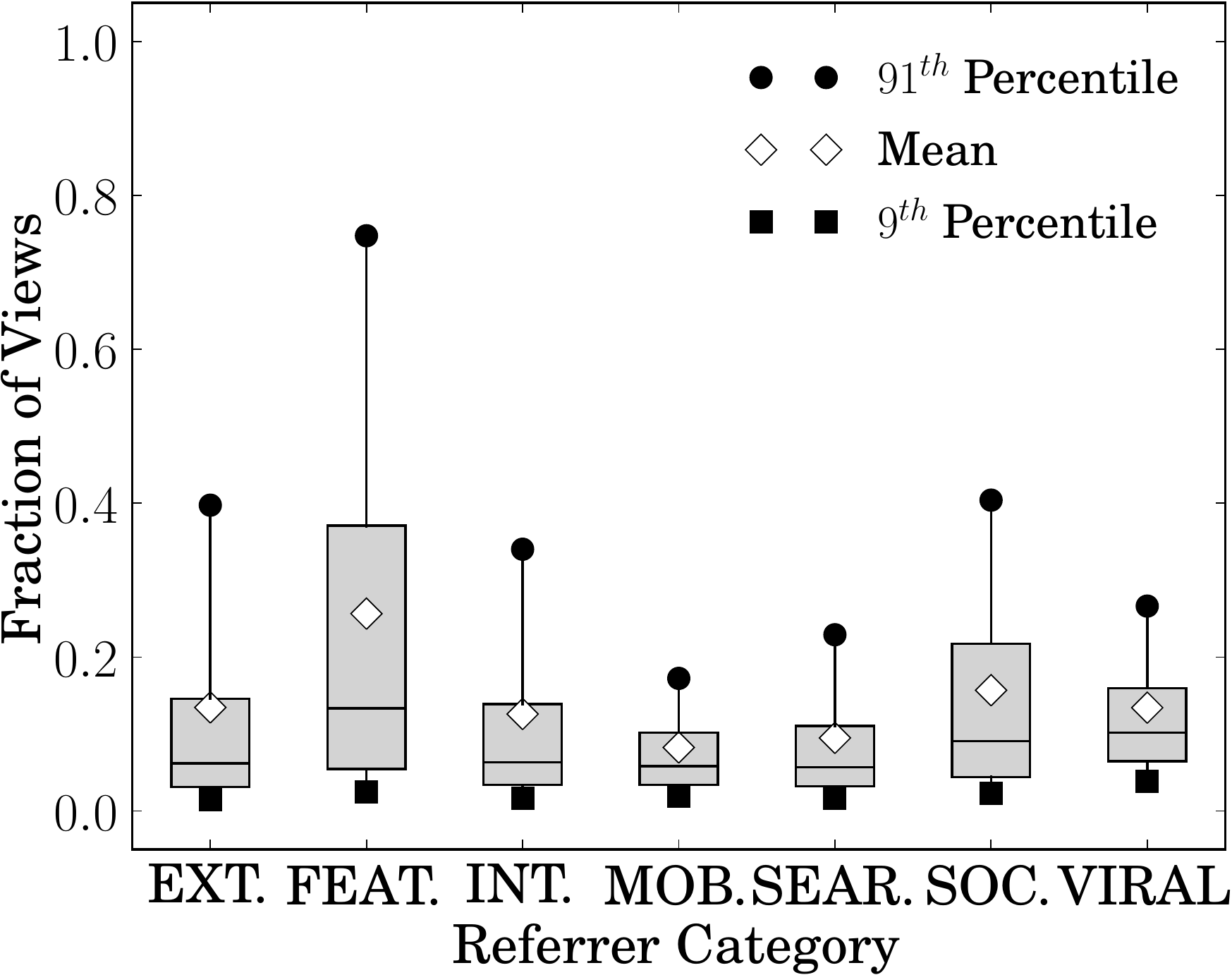}}}
 \mbox{\subfloat[Random]{\includegraphics[width=0.3\linewidth]{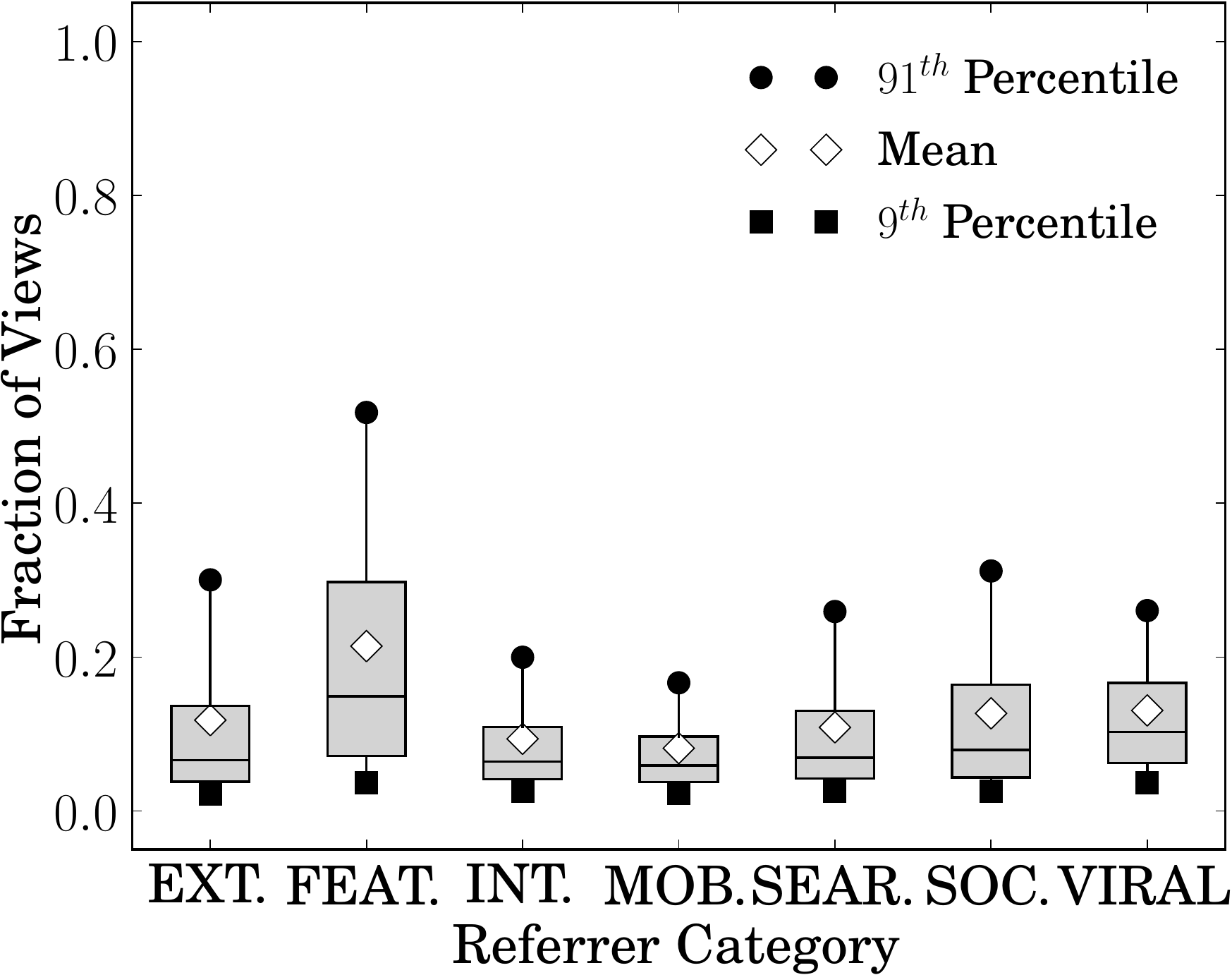}}}
 \caption{ Fraction of Views From Each Referrer Category.}
\vspace{-.9em}
 \label{fig:evview}
\end{figure}

We further analyze the importance of each referrer category by computing the distributions of the number of views for which each  category is responsible,  {\it taking only videos that received accesses from the given category}, and  computing percentages based on the  {\it  total views from referrers only (accounted views)}.  Figures \ref{fig:evview}(a-c) show
 box plots containing  $1^{st}$, $2^{nd}$ and $3^{rd}$ quartiles,  $9^{th}$ and $91^{th}$ percentiles, and the mean, for each  category and each  dataset\footnote{For any given referrer category, at least 1,000
videos received views for which it is responsible.}. Unlike Table \ref{tab:groups}, which shows aggregated results (i.e., results for {\it all videos in each dataset}), these plots allow us to assess the importance of each referrer category for individual videos. 

For example, Table \ref{tab:groups} shows that Social referrers do not appear to be  important for YouTomb dataset as a whole. However, taking only copyright  protected videos with at least one Social referrer, Figure \ref{fig:evview}-b) shows that, for 25\% of such videos ($1^{st}$ quartile),  more than 22\% of the accounted views come from subscription links. Thus, users do subscribe to other users who post copyright protected content. The Featured category is a similar case. 
For Top videos,  the Social, Featured and Viral categories are responsible for more than 30\%, 33\% and 34\%, respectively, of the accounted views for 25\% of the  videos with referrers from each such category (Figure \ref{fig:evview}-a). 
Finally,   Featured referrers play a key role to attract views to Random videos: 25\% of the videos with  Featured referrers received at least  30\% of the accounted views from such referrers (Figure \ref{fig:evview}-c).

\begin{figure}[t]
 \centering
 \mbox{\subfigure[Top]{\includegraphics[width=0.3\linewidth]{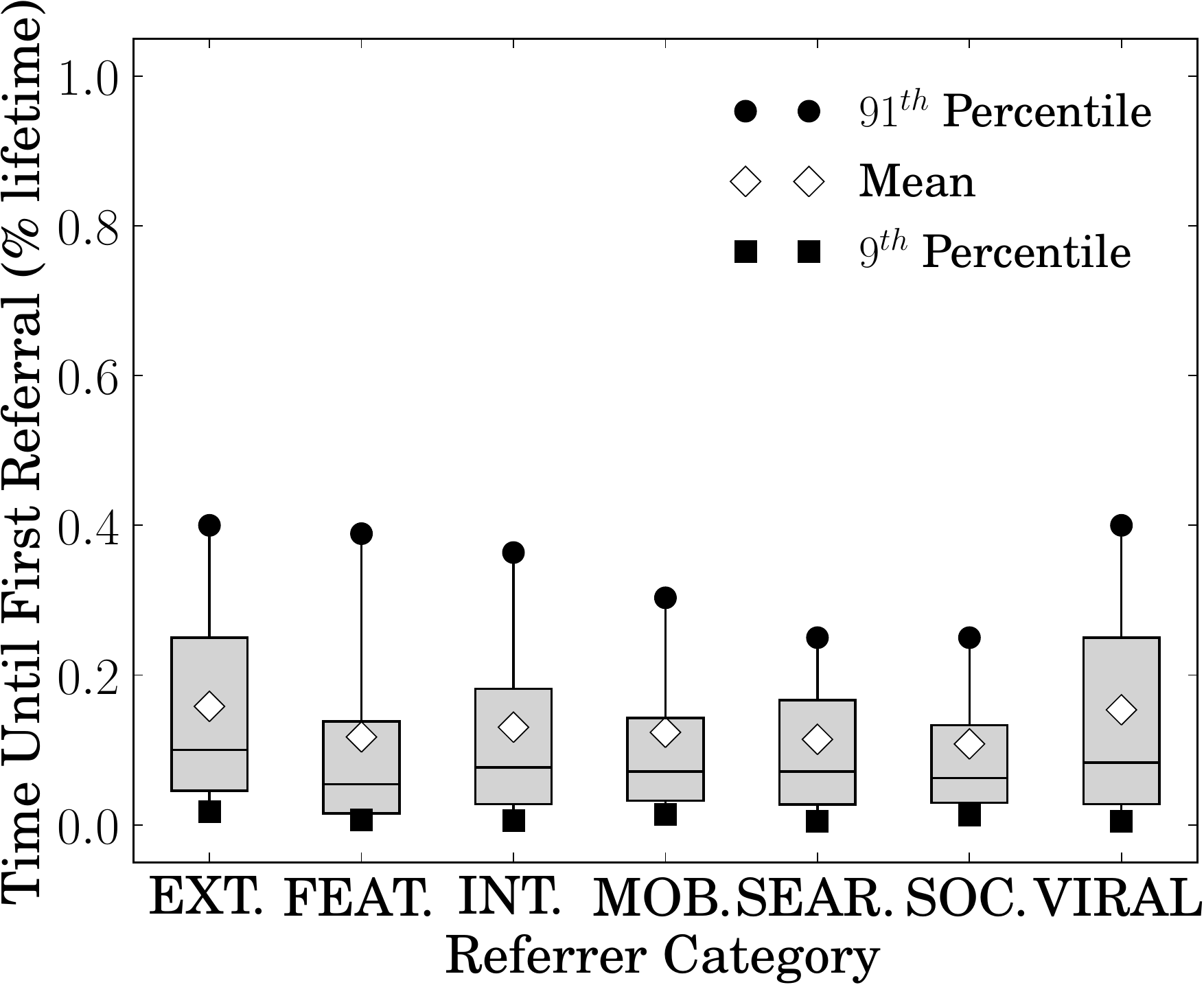}}}
 \mbox{\subfigure[YouTomb]{\includegraphics[width=0.3\linewidth]{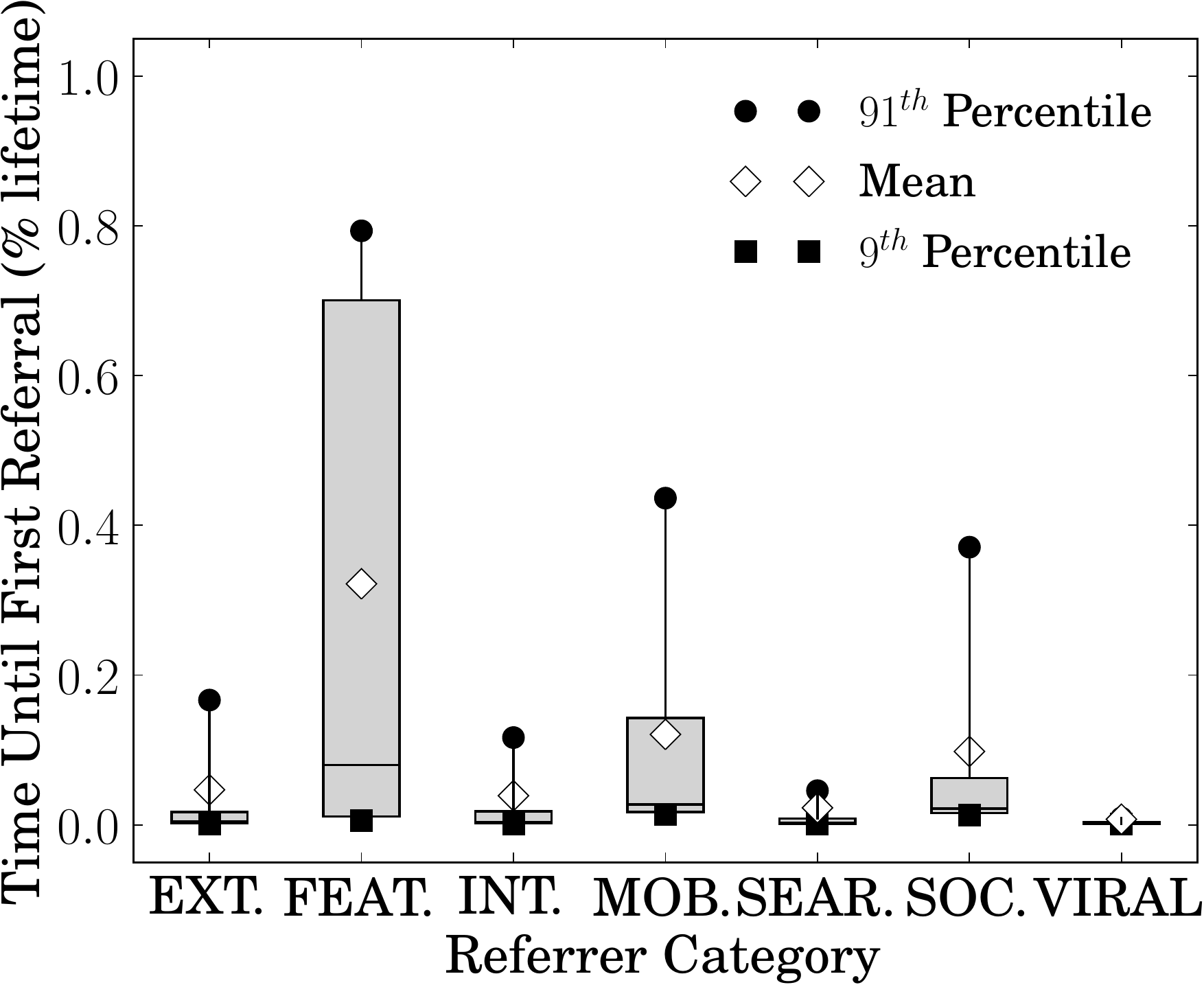}}}
 \mbox{\subfigure[Random]{\includegraphics[width=0.3\linewidth]{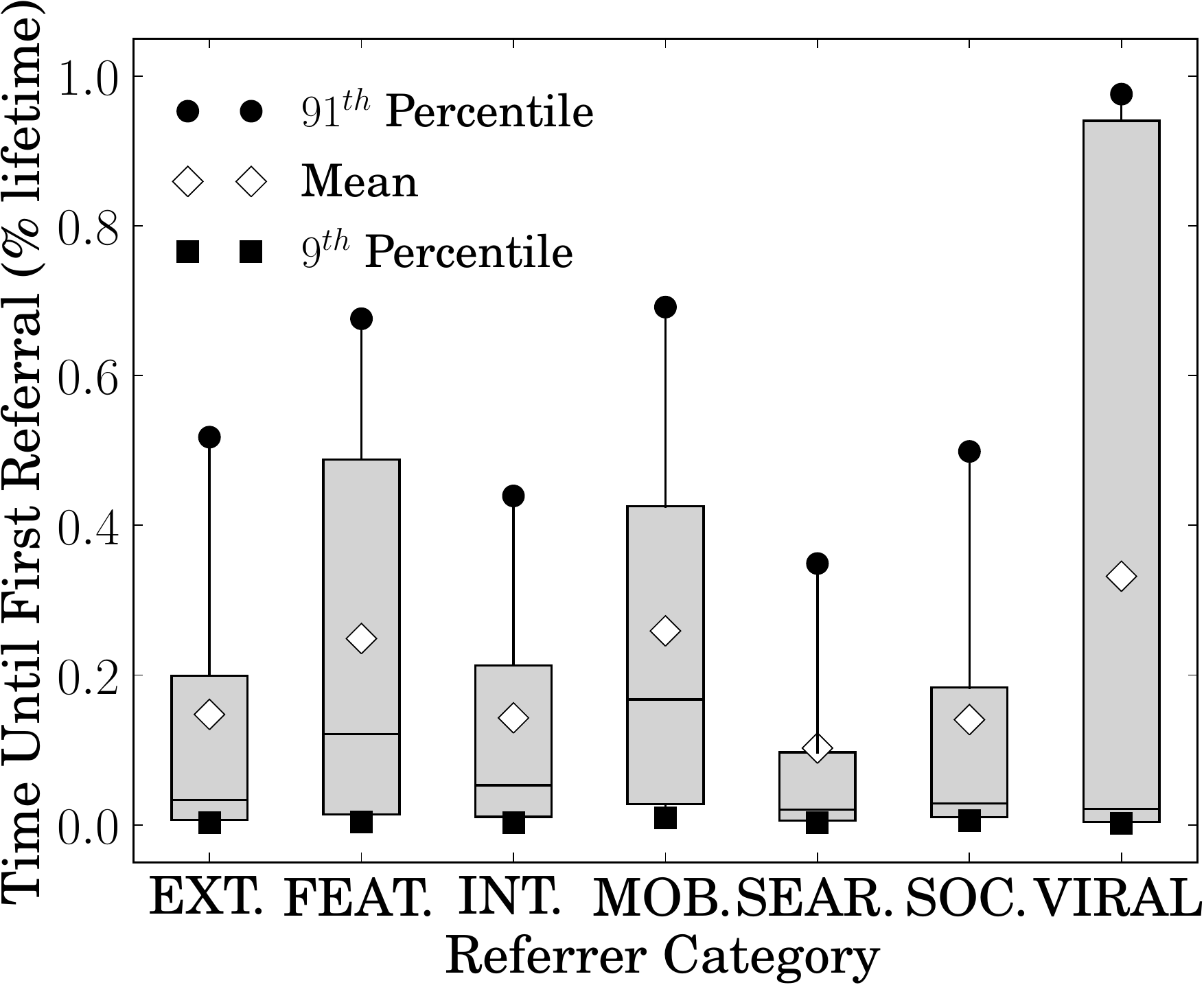}}}
 \caption{Time  Until the First Referrer Access (normalized by video's lifespan).}
 \label{fig:evdate}
\end{figure}

It is hard to tell whether one referrer  influences the number of views from other referrers. For example,  a Top video may experience a popularity growth from Social and Viral referrers {\it after} being featured in the top list. 
Next, we study this issue  by analyzing how early in a video's lifespan each type of referrer is used. 

\subsection{ How early do referrers appear (Q4b)?}  \label{subsec:Q4b}

We now analyze the referrers that first lead users to a video.  Table~\ref{tab:groups} also shows the fractions of videos that had the first referrer falling into each category ($f_{time}$). Since YouTube provides only the {\it day} each referrer was first used,  there might be ties with multiple categories, and the sum of  $f_{time}$  may exceed 100\% for a dataset. 

In general, viral spreading and internal YouTube mechanisms  appear as primary forms through which users reach the content for the first time, in all three datasets. For example, 
the first referrers for 79\%, 67\%, and 51\% of the Top videos are from the  Viral, Internal, and Mobile categories, respectively.  For the YouTomb dataset, Internal, Viral, and
Search contain the first referrers for 65\%, 62\% and 52\% of the videos, respectively.  For  the Random dataset, the first referrers of 55\%, 41\%, and 34\% of the videos are from the
Viral, Search, and Internal categories, respectively. 
Interestingly, mobile devices are also a relevant front door to Top videos, whereas for YouTomb and Random videos, the YouTube search engine accounts for a large fraction of the first
referrers.  

Figures \ref{fig:evdate}(a-c) show the distributions of the difference  between the time of the first referrer access and the time the video was uploaded, measured  as a fraction
of the video's lifespan. For the Top and YouTomb datasets, referrers (of any category) tend to happen very early: for 75\% of the Top and YouTomb videos, most referrer categories
have their first appearances during the first quarter of the video's lifespan. Indeed, only  9\% of the Top videos have their first referrer access (of any category) after  40\% of
their lifespans. The exception is the Featured category on YouTomb: those referrers tend to take more time to appear.  This suggests that YouTube may try to avoid featuring 
videos that are suspicious or have potential to be copyright protected.
For Random videos, in general, Search, Internal, External, and Social referrers tend to appear earlier than other types of referrers.  Thus, users are more likely to initially find
such videos through social links, search, other YouTube  mechanisms  or  external websites, instead of receiving them via e-mail or viewing them on mobile devices.

\subsection{Discussion}

We here focused on identifying the most important referrers that lead users to  a  video  (Q4). Our results are useful to help content creators to increase their viewership. For instance, search engines seem to attract most viewers to content, and they do so early on the video's  lifespan (Table~\ref{tab:groups}). However, focusing on particular videos, we find that  this may not hold for every case (Figure~\ref{fig:evview}). One  suggestion to content creators would thus be to provide good textual descriptions of video content, which would likely help search engine users to find it. Afterwards, a careful monitoring of how the video propagates on external websites and internal OSNs may be used to further boost viewership.

\section{Associations Between Various Features and Popularity (Q5)} \label{sec:rg1-corr}

We now tie the analyses of the previous sections together by assessing how different features are associated with the identified popularity trends, and also with total observed popularity values.  We first analyze whether videos that follow a similar popularity trend tend to have content in the same topic and  be reached through similar referrers (Section \ref{subsec:7a}). We then measure the correlations between various  features (shown in Table \ref{tab:feats}) and the popularity trend and observed popularity value of the videos (Section \ref{subsec:7b}). As in Section~\ref{sec:rg1-ksc}, we here focus on the Top and Random datasets.



\subsection{ What kinds of content and referrers are responsible for each popularity trend? (Q5a)} \label{subsec:7a}

\begin{figure}
 \centering
 \mbox{\subfigure[Top - Category]{\includegraphics[width=0.245\linewidth]{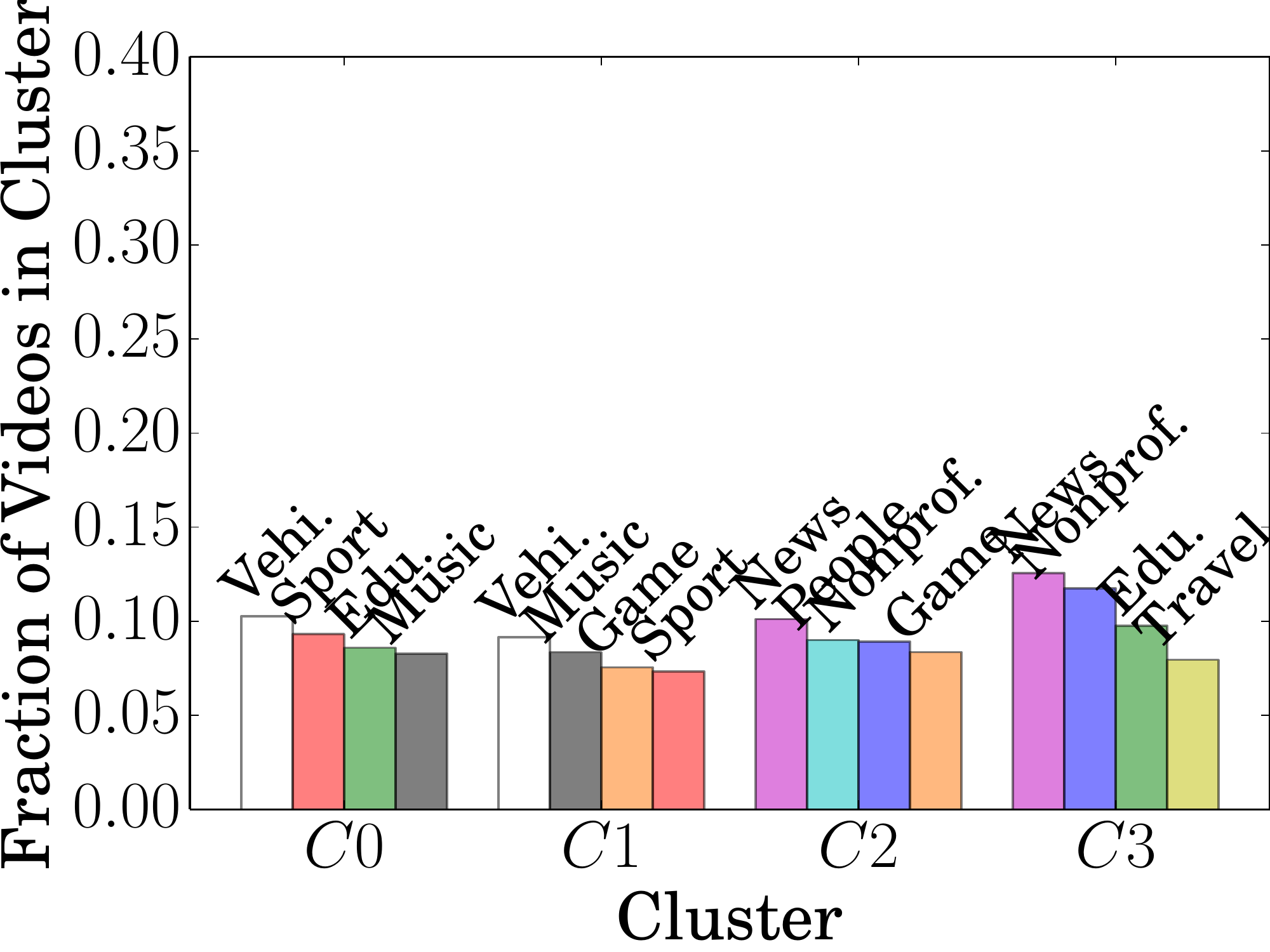}}}
 \mbox{\subfigure[Random - Category]{\includegraphics[width=0.245\linewidth]{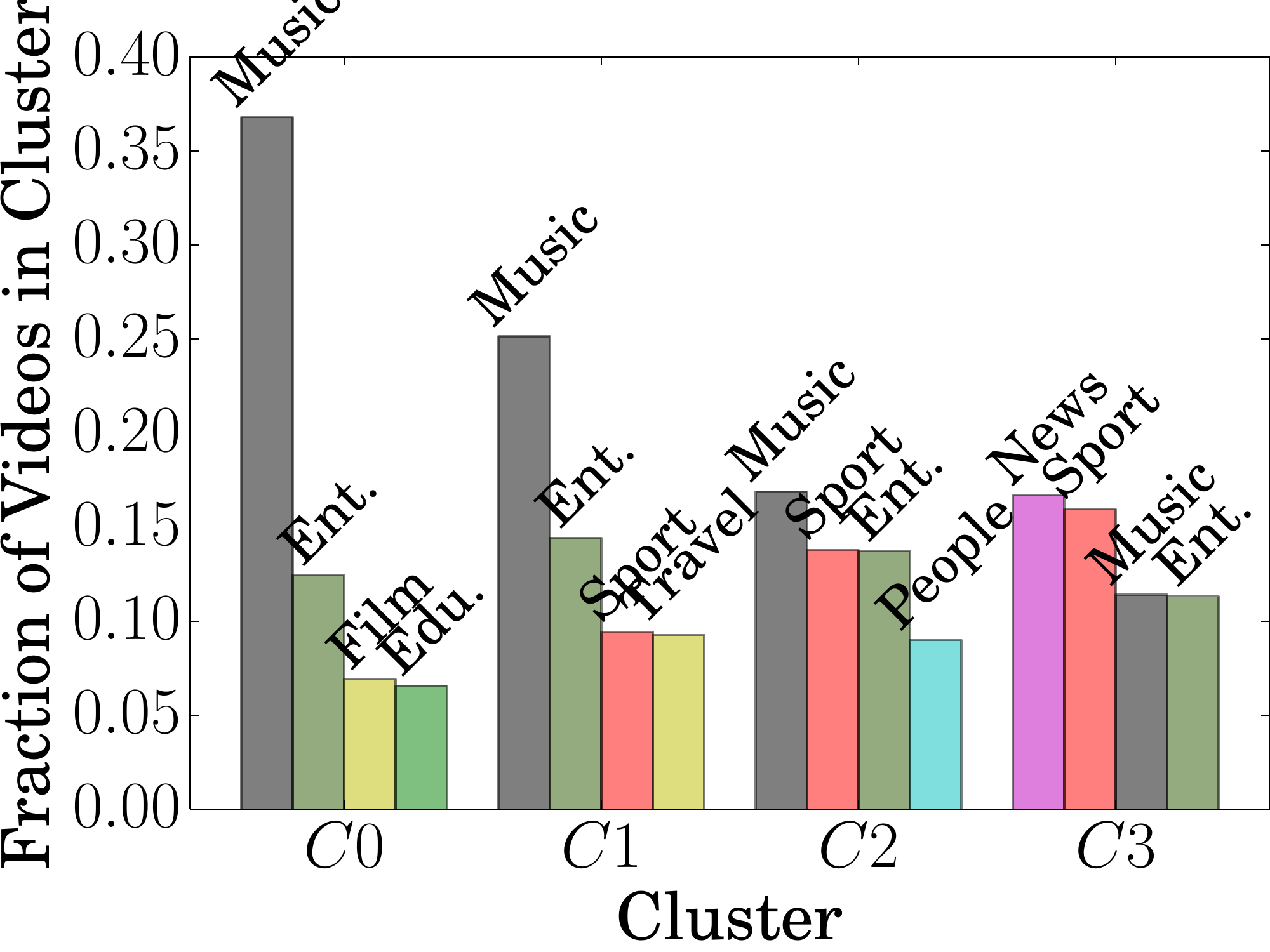}}}
 \mbox{\subfigure[Top - Referrer]{\includegraphics[width=0.245\linewidth]{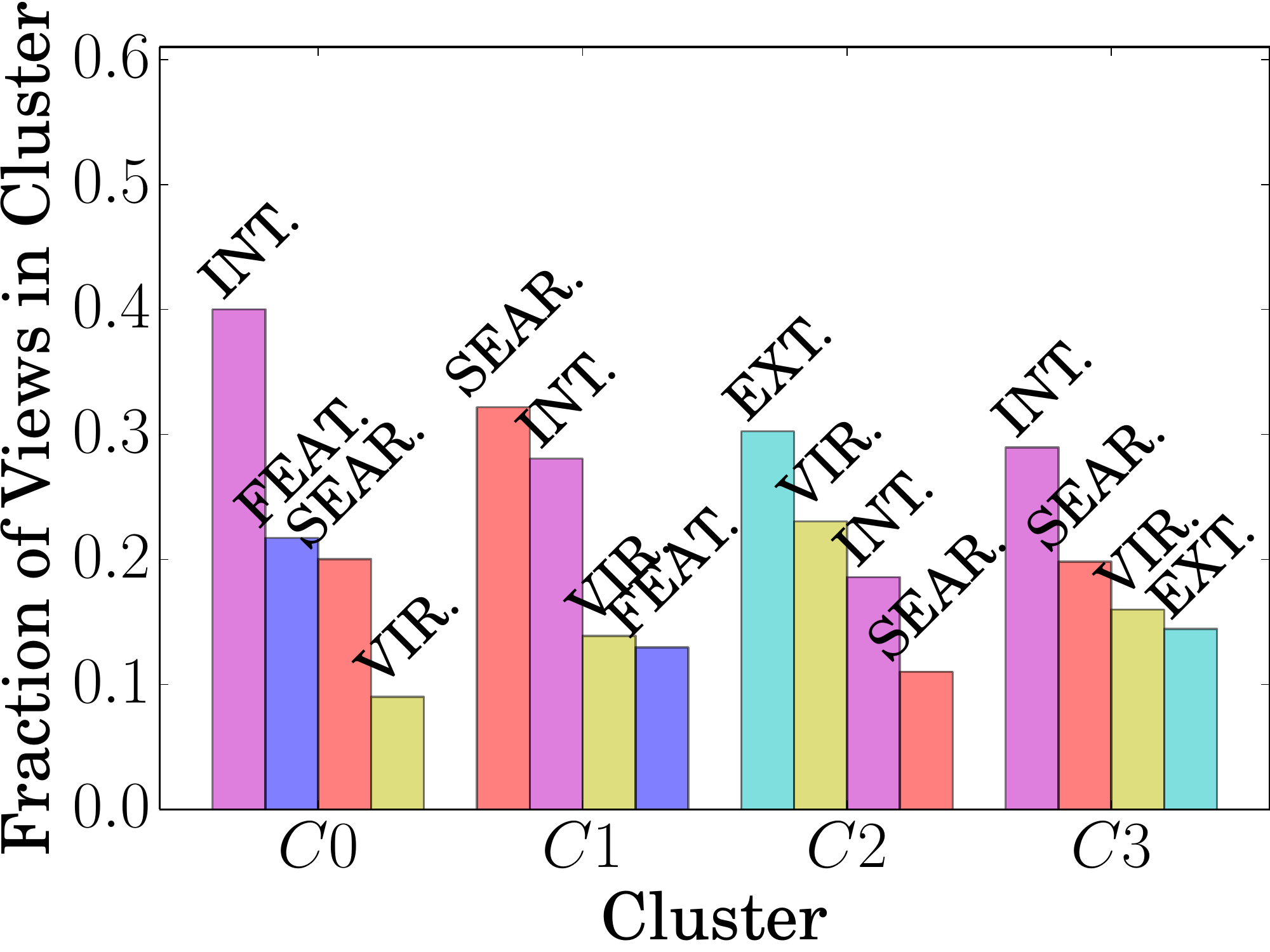}}}
 \mbox{\subfigure[Random - Referrer]{\includegraphics[width=0.245\linewidth]{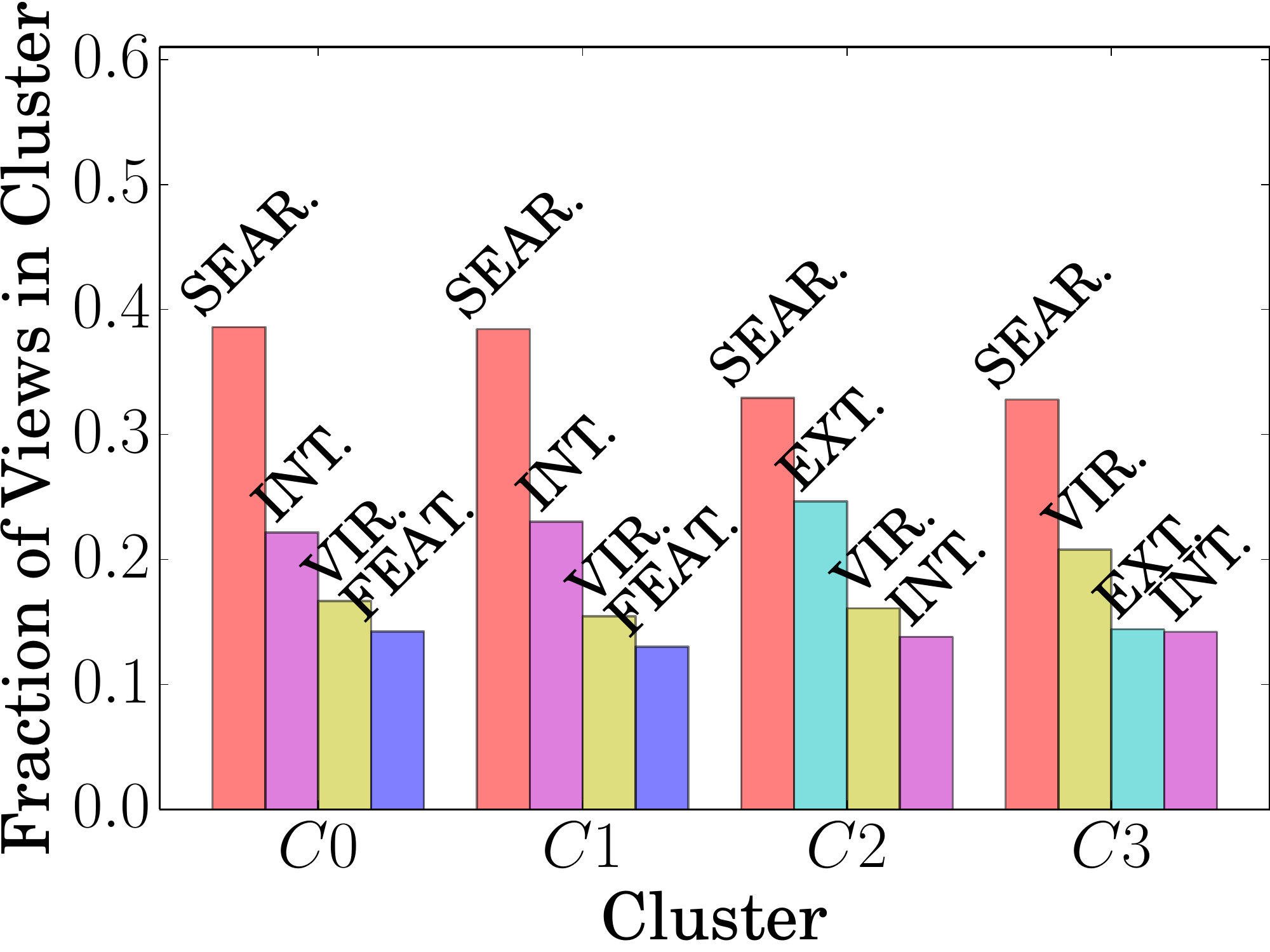}}}
 \caption{Fractions of Views Across Per Category (a,b) and Referrer (c,d) For Each Cluster.}
 \label{fig:cls-ch}
\end{figure}

We start by analyzing whether videos that
follow a similar popularity trend (same cluster) tend to have content in the video category.
For both datasets, we found that the distributions of the number of videos across categories in each cluster are statistically different from the distribution computed over all videos in the dataset, according to a Chi-Square test with p-value $< 0.01$.  Thus, videos in different clusters tend to be concentrated around different categories (or topics). In Figures \ref{fig:cls-ch}(a-b) we show the fractions of videos  in the top 4 categories in each cluster, for each dataset.



Starting with Top videos, Figure \ref{fig:cls-ch}(a) shows a clear divergence in the topics of the videos in each cluster. Clusters $C0$ and $C1$, which consist of videos that tend to attract viewers for longer periods, are composed mostly by videos about  music, sports, and automobiles, while  journalistic videos (news), video blogs (people) and videos related to activism  (non-profit) are the most common topics in clusters $C2$ and $C3$, which tend to  have  much shorter viewer retention periods.  This mostly likely occurs because such videos tend to be interesting only during short time periods.  For the Random dataset, Figure \ref{fig:cls-ch}(b) shows that videos with music and entertainment content are very frequent in all four clusters.  This may occur due to a natural bias of copyrighted content and of the queries used to build that dataset.  Regardless,  the frequencies of these categories tend to decrease, while news tends to become more frequent in clusters $C2$ and $C3$.

We now turn to the referrers used to reach videos in each cluster, and analyze the fractions of  views each type of referrer is responsible for, on average\footnote{These fractions are computed based on the total number of views  received through the referrers.}.  Once again,  for both datasets, the distributions of these fractions in each cluster are statistically different from the  distribution computed for all videos in the dataset. Thus, the types of referrers that attract the largest fractions of views do vary depending on the popularity trend. Figures \ref{fig:cls-ch}(c-d) show the results for each dataset, focusing again on the top 4 referrers per cluster.  Note that the only dataset where Search is the most important type of referrer for all trends is Random, due to the nature of its  crawling process. However,  Search, Internal and Viral referrers tend to be among the top 4 referrers in all  clusters of both  datasets.  Moreover, Featured referrers are among the most important ones for videos that remain attractive for some time ($C0$ and $C1$), while External referrers play an important role for videos that experience a sudden burst of popularity ($C2$ and $C3$).

\subsection{ What are the correlations between  features and popularity trends and values? (Q5b)}  \label{subsec:7b}

Finally, we measure the correlations between the features shown in Table \ref{tab:feats} and the popularity trends and the total popularity values of the videos (at the time of data crawling). To that end, we use
the maximal information coefficient, or MIC~\cite{Reshef2011}. MIC results range from 0, for no correlation,  to +1, for  strong (positive or negative) correlation.
This novel metric captures the normalized mutual information measure between two features. It  measures correlations
between different types of features (e.g., categorical and numeric) and is able to detect  non-linear and even periodic types of relationships, a limitation of other   coefficients  (e.g.,  Pearson and  Spearman). 
  We also used the Information Gain and the Gini coefficient~\cite{Cover1991} to measure the correlations, obtaining qualitatively similar results.

Since the values of some referrer and popularity features vary with time, we compute MIC results  for various {\it  monitoring periods}.  That is, we express the
monitoring period as a fraction of the video's lifespan, and compute feature values  only for that period. For example,  the correlation
between number of views and popularity trends for a monitoring period of 10\%  is computed taking  the number of views received during the first ten time
windows, since each time series has  100  windows.  By doing so, we can identify the most important features  in different phases of the video's lifespan. 

Table \ref{tab:mic} shows, for each dataset,  MIC
 results between the features and  {\it popularity trends}, for monitoring periods equal to 1\%, 5\%, 50\% and 100\%.  As the number of features is large,
we  aggregate MIC results for each feature class - video, referrer, and popularity, and  present mean ($\mu$) and maximum MIC for the
features in each class. 
Similarly, Table \ref{tab:mic-pop} shows the MIC results between features and  total  (observed)  {\it popularity values}.  Since  YouTube provides only the total number of views associated with each referrer, we only consider these features for a
monitoring period equal to 100\%, taking only the other referrer features (e.g., date of each referrer) for shorter periods.

\begin{table}[t]
\scriptsize
\centering
\caption{Average ($\mu$) and Maximum ($max$) Maximal Information Coefficient (MIC) Values per Feature Type for Popularity Trend.}
\begin{tabular}{llcccccccc} 
\toprule
&& \multicolumn{8}{c}{Monitoring Phase} \\
Dataset & Feature & \multicolumn{2}{c}{1\% of Lifespan} & \multicolumn{2}{c}{5\% of Lifespan} & \multicolumn{2}{c}{50\% of Lifespan} & \multicolumn{2}{c}{100\% of Lifespan} \\
\cmidrule(r){3-10}
&& $\mu$ & $max$ & $\mu$ & $max$ & $\mu$ & $max$ & $\mu$ & $max$ \\
\cmidrule(r){1-10}
\multirow{3}{*}{Top}             & Popularity Features       & .15 & .19 & .15 & .27 & .18 & .63 & .26 & .84\\
                                 & Referrer Features          & .11 & .17 & .11 & .18 & .12 & .18 & .11 & .19\\
                                 & Video Features         & .02 & .19 & .02 & .19 & .02 & .19 & .02 & .19\\ \midrule
                                 \midrule
\multirow{3}{*}{Random}          & Popularity Features       & .04 & .07 & .05 & .17 & .13 & .52 & .20 & .75\\
                                 & Referrer Features          & .04 & .08 & .04 & .08 & .05 & .08 & .08 & .15\\
                                 & Video Features         & .01 & .07 & .01 & .07 & .01 & .07 & .01 & .07\\
\bottomrule
\end{tabular}
\label{tab:mic}
\vspace{-1em}
\end{table}
\begin{table}[t]
\scriptsize
\centering
\caption{Average ($\mu$) and Maximum ($max$) Maximal Information Coefficient (MIC) Values per Feature Type for Total (Observed) Popularity Value.}
\begin{tabular}{llcccccccc}
\toprule
&& \multicolumn{8}{c}{Monitoring Phase} \\
Dataset & Feature & \multicolumn{2}{c}{1\% of Lifespan} & \multicolumn{2}{c}{5\% of Lifespan} & \multicolumn{2}{c}{50\% of Lifespan} & \multicolumn{2}{c}{100\% of Lifespan} \\
\cmidrule(r){3-10}
&& $\mu$ & $max$ & $\mu$ & $max$ & $\mu$ & $max$ & $\mu$ & $max$ \\
\cmidrule(r){1-10}
\multirow{3}{*}{Top}             & Popularity Features       & .31 & .48 & .32 & .56 & .41 & .88 & .57 & 1\\
                                 & Referrer Features          & .16 & .27 & .17 & .28 & .20 & .31 & .32 & .74\\
                                 & Video Features         & .11 & .31 & .11 & .31 & .11 & .31 & .11 & .31\\ \midrule
                                 \midrule
\multirow{3}{*}{Random}          & Popularity Features       & .16 & .32 & .22 & .48 & .36 & .89 & .51 & 1\\
                                 & Referrer Features          & .09 & .11 & .10 & .13 & .12 & .18 & .26 & .68\\
                                 & Video Features         & .08 & .18 & .08 & .18 & .08 & .18 & .08 & .18\\
\bottomrule
\end{tabular}
\label{tab:mic-pop}
\end{table}

We start by noting that, as the monitoring period increases, popularity features tend to greatly surpass the others
in importance for correlations with both  trends and popularity values. For trends, the popularity feature with maximum MIC  is peak fraction, whereas for popularity values, it is  number of views. For both of them, the
correlations are above  0.5   for monitoring periods beyond 50\%, in both datasets. However, for shorter periods,  the other (referrer and video) features  are also very important. This is interesting for popularity prediction tasks~\cite{Ahmed2013,Pinto2013,Radinsky2012,Szabo2010}, since  popularity features computed over short monitoring periods might be very unstable, and  prediction must rely mostly on other pieces of information about the video, such as its category and referrers.


\ch{
We further note that the relative average importance of each feature group is the same for both datasets: popularity features are more important than referrer features, which are more important than video features. 
The major differences between both datasets lie in the individual features within each feature class, as we discuss below. 

}

We now focus on the correlations computed for popularity trends (Table~\ref{tab:mic}). 
Albeit not shown in the table, for  short monitoring periods (e.g., 1\%), the most important video feature  is the video age (MIC  of 0.19 for Top and 0.07 for Random), while the number of views is the most important popularity feature
(MIC of 0.19 for Top  and 0.17 for Random).
The most important referrer feature  is the date of the first Internal referrer (MIC=0.17) for Top videos and the date of the first External referrer (MIC=0.08) for videos in  the Random dataset.
As time passes, the fraction of views on the peak window becomes the most important feature overall. This is expected, since
popularity trends are either concentrated on peaks or  exhibit  linear growth (Section \ref{sec:rg1-ksc}). 

In contrast, when correlating with the  total observed popularity values  (Table~\ref{tab:mic-pop}), the most important feature is  the number of views, for all  monitoring periods. At very early points in time (1\% of lifespan), the most important video 
 and referrer features  are video age (tied with upload date with MIC = 0.31) and the date of the first {\it Viral} referrer (MIC=0.26) for  Top videos. For the Random dataset,  they are video age (again, tied with upload date with MIC=0.18) and the date of the first  {\it Search} referrer (MIC=0.1). 

Thus, from the perspective of  popularity prediction, having fixed the monitoring period, the most important features to be explored depend on whether one aims at predicting a  trend or a value. For example, previous work showed that, by knowing the
trend of a video {\it before hand}, the accuracy of the prediction of popularity  values can be improved~\cite{Yang2010,Pinto2013}. However, we are not aware of any previous effort to predict the popularity  {\it trend} of a video (or UGC in general). As our results indicate, the use  of video and referrer  features can help in this task.

We also note that, when correlating with both trends and observed popularity values with a monitoring period of 1\%, the Music category is in the top 10 most correlated features for  the Random dataset\footnote{We represented each category by a binary feature, and computed correlations for each category separately.}.
The News category is also in the top 10 features when correlating with trends. 
This result is in agreement with  Figures~\ref{fig:cls-ch}(a-b), which show a more skewed concentration of categories across trends  in  the Random dataset.
Moreover,  when correlating with both trends and observed popularity values with a monitoring period of 100\%, some referrer features, mainly the number of views from the referrers, are in the top 10 most important features, in both datasets. We believe that such features would also be important at shorter monitoring periods. However,  their values are not available in our dataset. Thus we cannot test this hypothesis.



\subsection{Discussion}


The correlations unveiled in this section motivate the need to explore a diverse set of features for popularity prediction tasks. Most previous efforts explored only early points in the  popularity time
series~\cite{Ahmed2013,Pinto2013,Radinsky2012,Szabo2010}. Our results show that they could benefit from considering also other features. In particular, we found that while some  referrer and video features may be useful to predict, at very early stages in the video's lifespan, how its popularity will evolve over time (the trend), early popularity measures are the most useful features to  predict future popularity values. However, as discussed by previous work~\cite{Yang2010,Pinto2013}, one task may complement the other. \ch{Also, the differences  in relative importance of individual features  across datasets, particularly when considering early periods after video upload, raises a question of whether  different prediction methods (i.e., methods that exploit different sets of features) should be designed for different groups of videos. }

Other applications that may benefit from our results are recommender systems. By exploring important features that correlate with popularity, useful recommendations may be produced even before a video becomes popular. However, in this case a chicken-and-egg problem arises. Will a video become popular because it is  interesting or due to the recommendation engine? Investigating causality between factors that impact content popularity is  an important open question, which we leave for future work.

\section{Conclusions and Future Work} \label{sec:rg1-conc}

We have characterized the dynamics of video popularity on the currently most popular video sharing system, YouTube.  Driven by 5 research questions, we analyzed how the popularity of individual videos evolve since upload (Q1 and Q2), extracted  common trends of popularity evolution  (Q3), characterized the  types of referrers that  lead users to videos (Q4), and  correlated  popularity trends and final observed popularity values with various features (Q5). Our analyses were performed on three YouTube datasets, providing  a broad view of the popularity evolution for a diverse set of videos.

We found that copyright protected (YouTomb) videos tend to get most of their views much earlier in their lifespans, followed by Top videos, and then videos in the Random dataset.
We also found that Top videos tend to experience significant
popularity bursts, receiving a large fraction of their views on a
single day (or week). YouTomb videos also follow this pattern, and
this is less of a case for Random videos. However, using a time series clustering algorithm, we found that
the same 4 popularity trends seem to explain how video popularity evolves in both Top and Random datasets.

We also characterized the main referrers that led users to videos in each dataset. Particularly, we showed that  search and internal YouTube mechanisms, such as  lists of related videos,  are key mechanisms that attract users to the videos. Whereas Search referrers account for the largest fraction of views to videos in the Random dataset, internal  mechanisms play an even more important role to content dissemination for Top and YouTomb datasets. Also, our correlation results show that various video and referrer features can be explored for popularity prediction, and not only features extracted from early points of the popularity time series, as done by most  previous efforts. 

Our main findings can be applied in several contexts, as discussed next. \\

\vspace{-.8em}
\noindent \textbf{Content Distribution:} we found that, even after short monitoring periods, there exists some correlations between popularity trends and the analyzed features, motivating their use for predicting {\it popularity trends}.  Content distribution networks could use such predictions, together with observed popularity estimates, for load balancing, by provisioning videos predicted to remain popular for longer  (i.e., videos in $C0$ or $C1$) to more capable servers. For videos predicted to be in $C2$ or $C3$, as their popularity growth rates decrease there is a high chance that the attention for them will drop. Such videos should then be  provisioned by less capable servers or sent to secondary storage. Similarly, this knowledge could be used by ISPs for local caching.\\

\vspace{-.8em}
\noindent \textbf{Online Advertising:} our  results also suggest that different video categories tend to  more often follow different popularity trends (e.g., $C0$ and $C1$ are dominated by music and sports videos, while $C2$ and $C3$ by news and non-profit ones). This knowledge could be used by advertisers  to drive the selection of the video categories for ad placement, and by online advertising platforms to provide category-based price differentiation for advertisers (e.g., higher prices for categories that tend to remain popular for longer). Our results are also potentially useful for content publishers, who may profit from ads placed on their videos. The finding that Search  (and Featured) referrers attract more views for videos that remain attractive over time (i.e., videos in $C0$ and $C1$)  suggests that content publishers could periodically refine the keywords assigned to their videos (e.g.,  tags,
 title) to target different queries over time. For example, after a cycle of popularity growth and decay, publishers could adjust the video keywords and descriptions to possibly target other searchers that exploit related terms to find the video.\\

\vspace{-.8em}
\noindent \textbf{Monitoring Fame and Popularity:} From a social perspective, understanding content popularity could be used for monitoring fame and popularity of content producers, and  analyzing how users  seek-out and consume information on real world events  (e.g.,natural disasters, gossip news). As future work we aim at further investigating some of these issues, tackling questions such as: Is the consumption of content for different kinds of events  largely different? What are the most important blogs or personalities that drive attention to different events? How does content diffusion in one service, say Twitter or Facebook, impact the popularity of  videos on YouTube?  

\ch{ 
Finally,  although focused on YouTube videos, our work could be extended to tackle other types of content.
In particular, comparing how popularity evolves across different media types and the factors that are responsible for this evolution could be used by content producers and marketeers  to choose  the applications on which they should  focus. Another interesting direction for future work is the study of user popularity (as opposed to content popularity). Recent findings~\cite{Weng2012,Wattenhofer2012,Susarla2011} show that the amount of subscribers a user has plays a large role in the popularity of the content shared by her.  \ju{ We intend to extend our study to investigate the factors that impact user popularity on social media applications, as well as the inter dependencies that might exist between user and content popularity.} 
}


\bibliographystyle{ACM-Reference-Format-Journals}
\bibliography{bibs}


\begin{thebibliography}{00}


\ifx \showCODEN    \undefined \def \showCODEN     #1{\unskip}     \fi
\ifx \showDOI      \undefined \def \showDOI       #1{{\tt DOI:}\penalty0{#1}\ }
  \fi
\ifx \showISBNx    \undefined \def \showISBNx     #1{\unskip}     \fi
\ifx \showISBNxiii \undefined \def \showISBNxiii  #1{\unskip}     \fi
\ifx \showISSN     \undefined \def \showISSN      #1{\unskip}     \fi
\ifx \showLCCN     \undefined \def \showLCCN      #1{\unskip}     \fi
\ifx \shownote     \undefined \def \shownote      #1{#1}          \fi
\ifx \showarticletitle \undefined \def \showarticletitle #1{#1}   \fi
\ifx \showURL      \undefined \def \showURL       #1{#1}          \fi

\bibitem[\protect\citeauthoryear{Ahmed, Spagna, Huici, and Niccolini}{Ahmed
  et~al\mbox{.}}{2013}]%
        {Ahmed2013}
{Mohamed Ahmed}, {Stella Spagna}, {Felipe Huici}, {and} {Saverio Niccolini}.
  2013.
\newblock \showarticletitle{{A Peek Into the Future: Predicting the Evolution
  of Popularity in User Generated Content}}. In {\em Proc. WSDM}.
\newblock
\showISBNx{9781450318693}


\bibitem[\protect\citeauthoryear{Anderson, Kumar, Tomkins, and
  Vassilvitski}{Anderson et~al\mbox{.}}{2014}]%
        {Anderson}
{Ashton Anderson}, {Ravi Kumar}, {Andrew Tomkins}, {and} {Sergei Vassilvitski}.
  2014.
\newblock \showarticletitle{{Dynamics of Repeat Consumption}}. In {\em Proc.
  WWW}.
\newblock


\bibitem[\protect\citeauthoryear{Borghol, Ardon, Carlsson, Eager, and
  Mahanti}{Borghol et~al\mbox{.}}{2012}]%
        {Borghol2012}
{Youmna Borghol}, {Sebastien Ardon}, {Niklas Carlsson}, {Derek Eager}, {and}
  {Anirban Mahanti}. 2012.
\newblock \showarticletitle{{The Untold Story of the Clones: Content-agnostic
  Factors that Impact YouTube Video Popularity}}. In {\em Proc. KDD}.
\newblock
\showISBNx{9781450314626}


\bibitem[\protect\citeauthoryear{Borghol, Mitra, Ardon, Carlsson, Eager, and
  Mahanti}{Borghol et~al\mbox{.}}{2011}]%
        {Borghol2011}
{Youmna Borghol}, {Siddharth Mitra}, {Sebastien Ardon}, {Niklas Carlsson},
  {Derek Eager}, {and} {Anirban Mahanti}. 2011.
\newblock \showarticletitle{{Characterizing and Modeling Popularity of
  User-Generated Videos}}.
\newblock {\em Performance Evaluation\/} {68}, 11 (2011), 1037--1055.
\newblock
\showISSN{01665316}


\bibitem[\protect\citeauthoryear{Brodersen, Scellato, and
  Wattenhofer}{Brodersen et~al\mbox{.}}{2012}]%
        {Brodersen2012}
{Anders Brodersen}, {Salvatore Scellato}, {and} {Mirjam Wattenhofer}. 2012.
\newblock \showarticletitle{{YouTube Around the World}}. In {\em Proc. WWW}.
\newblock
\showISBNx{9781450312295}


\bibitem[\protect\citeauthoryear{Broxton, Interian, Vaver, and
  Wattenhofer}{Broxton et~al\mbox{.}}{2011}]%
        {Broxton2011}
{Tom Broxton}, {Yannet Interian}, {Jon Vaver}, {and} {Mirjam Wattenhofer}.
  2011.
\newblock \showarticletitle{{Catching a Viral Video}}.
\newblock {\em Journal of Intelligent Information Systems\/} (2011), 1--19.
\newblock
\showISSN{0925-9902}


\bibitem[\protect\citeauthoryear{Carrascosa, Gonz\'{a}lez, Cuevas, and
  Azcorra}{Carrascosa et~al\mbox{.}}{2013}]%
        {Carrascosa2013}
{Juan~Miguel Carrascosa}, {Roberto Gonz\'{a}lez}, {Rub\'{e}n Cuevas}, {and}
  {Arturo Azcorra}. 2013.
\newblock \showarticletitle{{Are trending topics useful for marketing?}}. In
  {\em Proc. COSN.}
\newblock
\showISBNx{9781450320849}


\bibitem[\protect\citeauthoryear{Cha, Benevenuto, Ahn, and Gummadi}{Cha
  et~al\mbox{.}}{2012}]%
        {Cha2012}
{Meeyoung Cha}, {Fabr\'{\i}cio Benevenuto}, {Yong-Yeol Ahn}, {and} {Krishna
  Gummadi}. 2012.
\newblock \showarticletitle{{Delayed information cascades in Flickr:
  Measurement, analysis, and modeling}}.
\newblock {\em Computer Networks\/} {56}, 3 (2012), 1066--1076.
\newblock
\showISSN{13891286}


\bibitem[\protect\citeauthoryear{Cha, Kwak, Rodriguez, Ahn, and Moon}{Cha
  et~al\mbox{.}}{2009}]%
        {Cha2009}
{Meeyoung Cha}, {Haewoon Kwak}, {Pablo Rodriguez}, {Yong-Yeol Ahn}, {and} {Sue
  Moon}. 2009.
\newblock \showarticletitle{{Analyzing the Video Popularity Characteristics of
  Large-Scale User Generated Content Systems}}.
\newblock {\em IEEE/ACM Transactions on Networking\/} {17}, 5 (2009),
  1357--1370.
\newblock
\showISSN{1063-6692}


\bibitem[\protect\citeauthoryear{Chatzopoulou, Sheng, and
  Faloutsos}{Chatzopoulou et~al\mbox{.}}{2009}]%
        {Chatzopoulou}
{Gloria Chatzopoulou}, {Cheng Sheng}, {and} {Michalis Faloutsos}. 2009.
\newblock \showarticletitle{{A First Step Towards Understanding Popularity in
  YouTube}}. In {\em Proc. Infocom Workshops.}
\newblock
\showISBNx{978-1-4244-6739-6}


\bibitem[\protect\citeauthoryear{Clauset, Shalizi, and Newman}{Clauset
  et~al\mbox{.}}{2009}]%
        {Clauset2009}
{Aaron Clauset}, {Cosma~Rohilla Shalizi}, {and} {M.~E.~J. Newman}. 2009.
\newblock \showarticletitle{{Power-Law Distributions in Empirical Data}}.
\newblock {\it SIAM Rev.} {51}, 4 (2009), 661--703.
\newblock
\showISSN{0036-1445}


\bibitem[\protect\citeauthoryear{Conover, Ferrara, Menczer, and
  Flammini}{Conover et~al\mbox{.}}{2013}]%
        {Conover2013}
{Michael~D Conover}, {Emilio Ferrara}, {Filippo Menczer}, {and} {Alessandro
  Flammini}. 2013.
\newblock \showarticletitle{{The digital evolution of occupy wall street.}}
\newblock {\em PloS one\/} {8}, 5 (2013), e64679.
\newblock
\showISSN{1932-6203}


\bibitem[\protect\citeauthoryear{Cormode and Krishnamurthy}{Cormode and
  Krishnamurthy}{2008}]%
        {Cormode2008}
{Graham Cormode} {and} {Balachander Krishnamurthy}. 2008.
\newblock \showarticletitle{{Key Differences Between Web1.0 and Web2.0}}.
\newblock {\em First Monday\/} {13}, 6 (2008).
\newblock


\bibitem[\protect\citeauthoryear{Cover and Thomas}{Cover and Thomas}{2006}]%
        {Cover1991}
{Thomas~M Cover} {and} {Joy~A Thomas}. 2006.
\newblock {\em {Elements of Information Theory}}. Vol.~6.
\newblock Wiley.
\newblock


\bibitem[\protect\citeauthoryear{Crane and Sornette}{Crane and
  Sornette}{2008}]%
        {Crane2008}
{Riley Crane} {and} {Didier Sornette}. 2008.
\newblock \showarticletitle{{Robust Dynamic Classes Revealed by Measuring the
  Response Function of a Social System}}.
\newblock {\em Proceedings of the National Academy of Sciences\/} {105}, 41
  (2008), 15649--53.
\newblock
\showISSN{1091-6490}


\bibitem[\protect\citeauthoryear{Easley and Kleinberg}{Easley and
  Kleinberg}{2010}]%
        {Easley2010}
{David Easley} {and} {Jon Kleinberg}. 2010.
\newblock {\em {Networks, Crowds, and Markets: Reasoning About a Highly
  Connected World}\/} (1 ed.).
\newblock Cambridge University Press.
\newblock
\showISBNx{0521195330, 9780521195331}


\bibitem[\protect\citeauthoryear{Ferrara, Varol, Menczer, and Flammini}{Ferrara
  et~al\mbox{.}}{2013}]%
        {Ferrara2013}
{Emilio Ferrara}, {Onur Varol}, {Filippo Menczer}, {and} {Alessandro Flammini}.
  2013.
\newblock \showarticletitle{{Traveling trends: social butterflies or frequent
  fliers?}}. In {\em Proc. COSN}.
\newblock
\showISBNx{9781450320849}


\bibitem[\protect\citeauthoryear{Figueiredo, Benevenuto, and
  Almeida}{Figueiredo et~al\mbox{.}}{2011}]%
        {Figueiredo2011}
{Flavio Figueiredo}, {Fabr\'{\i}cio Benevenuto}, {and} {Jussara Almeida}. 2011.
\newblock \showarticletitle{{The Tube Over Time: Characterizing Popularity
  Growth of YouTube Videos}}. In {\em Proc. WSDM}.
\newblock
\showISBNx{9781450304931}


\bibitem[\protect\citeauthoryear{Figueiredo, Pinto, Bel\'{e}m, Almeida,
  Gon\c{c}alves, Fernandes, and Moura}{Figueiredo et~al\mbox{.}}{2012}]%
        {Figueiredo2012}
{Flavio Figueiredo}, {Henrique Pinto}, {Fabiano Bel\'{e}m}, {Jussara Almeida},
  {Marcos Gon\c{c}alves}, {David Fernandes}, {and} {Edleno Moura}. 2012.
\newblock \showarticletitle{{Assessing the Quality of Textual Features in
  Social Media}}.
\newblock {\em Information Processing \& Management\/} (2012).
\newblock
\showISSN{03064573}


\bibitem[\protect\citeauthoryear{Golder and Hubberman}{Golder and
  Hubberman}{2006}]%
        {Golder2006}
{Scott~A. Golder} {and} {Bernado Hubberman}. 2006.
\newblock \showarticletitle{{Usage Patterns of Collaborative Tagging Systems}}.
\newblock {\em Journal of Information Science\/} {32}, 2 (2006), 198--208.
\newblock
\showISSN{0165-5515}


\bibitem[\protect\citeauthoryear{Islam, Eager, Carlsson, and Mahanti}{Islam
  et~al\mbox{.}}{2013}]%
        {Islam2013}
{M.~Aminul Islam}, {Derek Eager}, {Niklas Carlsson}, {and} {Anirban Mahanti}.
  2013.
\newblock \showarticletitle{{Revisiting Popularity Characterization and
  Modeling of User-generated Videos}}. In {\em Proc. Mascots}.
\newblock


\bibitem[\protect\citeauthoryear{Jiang, Miao, Yang, Lan, and Hauptmann}{Jiang
  et~al\mbox{.}}{2014}]%
        {Jiang2014}
{Lu Jiang}, {Yajie Miao}, {Yi Yang}, {Zhenzhong Lan}, {and} {Alexander~G.
  Hauptmann}. 2014.
\newblock \showarticletitle{{Viral Video Style: A Closer Look at Viral Videos
  on YouTube}}. In {\em Proc. ICMR}.
\newblock


\bibitem[\protect\citeauthoryear{Kamath, Caverlee, Lee, and Cheng}{Kamath
  et~al\mbox{.}}{2013}]%
        {Kamath2013}
{Krishna~Y. Kamath}, {James Caverlee}, {Kyumin Lee}, {and} {Zhiyuan Cheng}.
  2013.
\newblock \showarticletitle{{Spatio-temporal dynamics of online memes: a study
  of geo-tagged tweets}}. In {\em Proc WWW.}
\newblock
\showISBNx{978-1-4503-2035-1}


\bibitem[\protect\citeauthoryear{Kaplan and Haenlein}{Kaplan and
  Haenlein}{2010}]%
        {Kaplan2010}
{Andreas~M. Kaplan} {and} {Michael Haenlein}. 2010.
\newblock \showarticletitle{{Users of the World, Unite! The Challenges and
  Opportunities of Social Media}}.
\newblock {\em Business Horizons\/} {53}, 1 (2010), 59--68.
\newblock
\showISSN{00076813}


\bibitem[\protect\citeauthoryear{Kasneci, Ramanath, Suchanek, and
  Weikum}{Kasneci et~al\mbox{.}}{2009}]%
        {Kasneci2009}
{Gjergji Kasneci}, {Maya Ramanath}, {Fabian Suchanek}, {and} {Gerhard Weikum}.
  2009.
\newblock \showarticletitle{{The YAGO-NAGA approach to knowledge discovery}}.
\newblock {\em ACM SIGMOD Record\/} {37}, 4 (2009), 41.
\newblock
\showISSN{01635808}


\bibitem[\protect\citeauthoryear{Khosla, Sarma, and Hamid}{Khosla
  et~al\mbox{.}}{2014}]%
        {Khosla2014}
{Aditya Khosla}, {Atish~Das Sarma}, {and} {Raffay Hamid}. 2014.
\newblock \showarticletitle{{What Makes an Image Popular}}. In {\em Proc. WWW}.
\newblock


\bibitem[\protect\citeauthoryear{Lakkaraju, McAuley, and Leskovec}{Lakkaraju
  et~al\mbox{.}}{2013}]%
        {Lakkaraju2013}
{Himabindu Lakkaraju}, {Julian McAuley}, {and} {Jure Leskovec}. 2013.
\newblock \showarticletitle{{What's in a Name? Understanding the Interplay
  between Titles, Content, and Communities in Social Media}}. In {\em Proc.
  ICWSM}.
\newblock


\bibitem[\protect\citeauthoryear{Lehmann, Gon\c{c}alves, Ramasco, and
  Cattuto}{Lehmann et~al\mbox{.}}{2012}]%
        {Lehmann2012}
{Janette Lehmann}, {Bruno Gon\c{c}alves}, {Jos\'{e}~J. Ramasco}, {and} {Ciro
  Cattuto}. 2012.
\newblock \showarticletitle{{Dynamical classes of collective attention in
  twitter}}. In {\em Proc. WWW}.
\newblock
\showISBNx{9781450312295}


\bibitem[\protect\citeauthoryear{Lerman and Jones}{Lerman and Jones}{2006}]%
        {Lerman2006}
{Kristina Lerman} {and} {Laurie Jones}. 2006.
\newblock \showarticletitle{{Social Browsing on Flickr}}. In {\em Proc. ICWSM}.
\newblock


\bibitem[\protect\citeauthoryear{Li, Ma, Wang, Liu, and Xu}{Li
  et~al\mbox{.}}{2013}]%
        {Li2013}
{Haitao Li}, {Xiaoqiang Ma}, {Feng Wang}, {Jiangchuan Liu}, {and} {Ke Xu}.
  2013.
\newblock \showarticletitle{{On popularity prediction of videos shared in
  online social networks}}. In {\em Proc. CIKM}.
\newblock


\bibitem[\protect\citeauthoryear{Lin, Keogh, Wei, and Lonardi}{Lin
  et~al\mbox{.}}{2007}]%
        {Lin2007}
{Jessica Lin}, {Eamonn Keogh}, {Li Wei}, {and} {Stefano Lonardi}. 2007.
\newblock \showarticletitle{{Experiencing SAX: A Novel Symbolic representation
  of Time Series}}.
\newblock {\em Data Mining and Knowledge Discovery\/}  {15} (2007), 107--144.
\newblock
Issue 2.


\bibitem[\protect\citeauthoryear{Marlow, Naaman, Boyd, and Davis}{Marlow
  et~al\mbox{.}}{2006}]%
        {Marlow2006}
{Cameron Marlow}, {Mor Naaman}, {Danah Boyd}, {and} {Marc Davis}. 2006.
\newblock \showarticletitle{{HT06, tagging paper, taxonomy, Flickr, academic
  article, to read}}. In {\em Prog. HT}.
\newblock
\showISBNx{1595934170}


\bibitem[\protect\citeauthoryear{Matsubara, Sakurai, Prakash, Li, and
  Faloutsos}{Matsubara et~al\mbox{.}}{2012}]%
        {Matsubara2012}
{Yasuko Matsubara}, {Yasushi Sakurai}, {B.~Aditya Prakash}, {Lei Li}, {and}
  {Christos Faloutsos}. 2012.
\newblock \showarticletitle{{Rise and Fall Patterns of Information Diffusion}}.
  In {\em Proc. KDD}.
\newblock
\showISBNx{9781450314626}


\bibitem[\protect\citeauthoryear{Menasc\'{e} and Almeida}{Menasc\'{e} and
  Almeida}{2002}]%
        {Menasce2002}
{Daniel Menasc\'{e}} {and} {Virgilio Almeida}. 2002.
\newblock {\em {Capacity Planning for Web Services: Metrics, Models, and
  Methods}}.
\newblock Prentice Hall.
\newblock


\bibitem[\protect\citeauthoryear{Mesty\'{a}n, Yasseri, and
  Kert\'{e}sz}{Mesty\'{a}n et~al\mbox{.}}{2013}]%
        {Mestyan2013}
{M\'{a}rton Mesty\'{a}n}, {Taha Yasseri}, {and} {J\'{a}nos Kert\'{e}sz}. 2013.
\newblock \showarticletitle{{Early prediction of movie box office success based
  on Wikipedia activity big data.}}
\newblock {\em PloS one\/} {8}, 8 (2013), e71226.
\newblock
\showISSN{1932-6203}


\bibitem[\protect\citeauthoryear{Moat, Curme, Avakian, Kenett, Stanley, and
  Preis}{Moat et~al\mbox{.}}{2013}]%
        {Moat2013}
{Helen~Susannah Moat}, {Chester Curme}, {Adam Avakian}, {Dror Kenett}, {Eugene
  Stanley}, {and} {Tobias Preis}. 2013.
\newblock \showarticletitle{{Quantifying Wikipedia Usage Patterns Before Stock
  Market Moves}}.
\newblock {\em Scientific Reports\/}  {3} (2013).
\newblock
\showISSN{2045-2322}


\bibitem[\protect\citeauthoryear{Oliveira, Cherubini, and Oliver}{Oliveira
  et~al\mbox{.}}{2010}]%
        {Oliveira2010}
{Rodrigo~De Oliveira}, {Mauro Cherubini}, {and} {Nuria Oliver}. 2010.
\newblock \showarticletitle{{Looking at Near-Duplicate Videos from a
  Human-Centric Perspective}}.
\newblock {\em ACM Transactions on Multimedia Computing, Communications, and
  Applications\/} {6}, 3 (2010), 1--22.
\newblock
\showISSN{15516857}


\bibitem[\protect\citeauthoryear{Pinto, Almeida, and Gon\c{c}alves}{Pinto
  et~al\mbox{.}}{2013}]%
        {Pinto2013}
{Henrique Pinto}, {Jussara Almeida}, {and} {Marcos Gon\c{c}alves}. 2013.
\newblock \showarticletitle{{Using Early View Patterns to Predict the
  Popularity of YouTube Videos}}. In {\em Proc. WSDM}.
\newblock


\bibitem[\protect\citeauthoryear{Preis, Moat, and Stanley}{Preis
  et~al\mbox{.}}{2013}]%
        {Preis2013}
{Tobias Preis}, {Helen~Susannah Moat}, {and} {H~Eugene Stanley}. 2013.
\newblock \showarticletitle{{Quantifying trading behavior in financial markets
  using Google Trends.}}
\newblock {\em Scientific reports\/}  {3} (2013), 1684.
\newblock
\showISSN{2045-2322}


\bibitem[\protect\citeauthoryear{Radinsky, Svore, Dumais, Teevan, Bocharov, and
  Horvitz}{Radinsky et~al\mbox{.}}{2013}]%
        {Radinsky2012}
{Kira Radinsky}, {Krysta Svore}, {Susan Dumais}, {Jaime Teevan}, {Alex
  Bocharov}, {and} {Eric Horvitz}. 2013.
\newblock \showarticletitle{{Behavioral Dynamics on the Web: Learning,
  Modeling, and Prediction}}.
\newblock {\em ACM Transactions on Information Systems\/} {32}, 3 (2013),
  1--37.
\newblock
\showISBNx{9781450312295}


\bibitem[\protect\citeauthoryear{Ratkiewicz, Flammini, and Menczer}{Ratkiewicz
  et~al\mbox{.}}{2010}]%
        {Ratkiewicz2010}
{Jacob Ratkiewicz}, {Alessandro Flammini}, {and} {Fillipo Menczer}. 2010.
\newblock \showarticletitle{{Traffic in social media I: paths through
  information networks}}. In {\em Proc. SIN}.
\newblock


\bibitem[\protect\citeauthoryear{Reshef, Reshef, Finucane, {Sharon R.
  Grossman}, McVean, Turnbaugh, Lander, Mitzenmacher, and Sabeti}{Reshef
  et~al\mbox{.}}{2011}]%
        {Reshef2011}
{David Reshef}, {Yakir Reshef}, {Hilary~K. Finucane}, {{Sharon R. Grossman}},
  {Gilean McVean}, {Peter~J. Turnbaugh}, {Eric~S. Lander}, {Michael
  Mitzenmacher}, {and} {Pardis~C. Sabeti}. 2011.
\newblock \showarticletitle{{Detecting Novel Associations in Large Data Sets}}.
\newblock {\em Science\/} {334}, 6062 (2011), 1518--1524.
\newblock
\showISSN{00368075}


\bibitem[\protect\citeauthoryear{Sinha and Pan}{Sinha and Pan}{2007}]%
        {Sinha2007}
{Sitabhra Sinha} {and} {Raj~Kumar Pan}. 2007.
\newblock \showarticletitle{{How a "Hit" is Born: The Emergence of Popularity
  from the Dynamics of Collective Choice}}.
\newblock In {\em Econophysics and Sociophysics: Trends and Perspectives}.
  Wiley, Online.
\newblock


\bibitem[\protect\citeauthoryear{Susarla, Oh, and Tan}{Susarla
  et~al\mbox{.}}{2011}]%
        {Susarla2011}
{Anjana Susarla}, {Jeong-Ha Oh}, {and} {Yong Tan}. 2011.
\newblock \showarticletitle{{Social Networks and the Diffusion of
  User-Generated Content: Evidence from YouTube}}.
\newblock {\em Information Systems Research\/} {23}, 1 (2011), 1--19.
\newblock
\showISSN{10477047}


\bibitem[\protect\citeauthoryear{Szabo and Huberman}{Szabo and
  Huberman}{2010}]%
        {Szabo2010}
{Gabor Szabo} {and} {Bernardo Huberman}. 2010.
\newblock \showarticletitle{{Predicting the Popularity of Online Content}}.
\newblock {\it Commun. ACM} {53}, 8 (2010), 80--88.
\newblock
\showISSN{1556-5068}


\bibitem[\protect\citeauthoryear{Vakali, Giatsoglou, and Antaris}{Vakali
  et~al\mbox{.}}{2012}]%
        {Vakali2012}
{Athena Vakali}, {Maria Giatsoglou}, {and} {Stefanos Antaris}. 2012.
\newblock \showarticletitle{{Social networking trends and dynamics detection
  via a cloud-based framework design}}. In {\em Proc. WWW}.
\newblock
\showISBNx{9781450312301}


\bibitem[\protect\citeauthoryear{van Zwol}{van Zwol}{2007}]%
        {Zwol2007}
{Roelof van Zwol}. 2007.
\newblock \showarticletitle{{Flickr: Who is Looking?}}. In {\em Proc. WI}.
  IEEE.
\newblock
\showISBNx{0-7695-3026-5}


\bibitem[\protect\citeauthoryear{Wang and Huberman}{Wang and Huberman}{2012}]%
        {Wang2012}
{Chunyan Wang} {and} {Bernardo Huberman}. 2012.
\newblock \showarticletitle{{Long trend dynamics in social media}}.
\newblock {\em EPJ Data Science\/} {1}, 2 (2012).
\newblock


\bibitem[\protect\citeauthoryear{Wattenhofer, Wattenhofer, and Zhu}{Wattenhofer
  et~al\mbox{.}}{2012}]%
        {Wattenhofer2012}
{Mirjam Wattenhofer}, {Roger Wattenhofer}, {and} {Zack Zhu}. 2012.
\newblock \showarticletitle{{The YouTube social network}}. In {\em Proc.
  ICWSM}.
\newblock


\bibitem[\protect\citeauthoryear{Weng, Flammini, Vespignani, and Menczer}{Weng
  et~al\mbox{.}}{2012}]%
        {Weng2012}
{Lillian Weng}, {Alessandro Flammini}, {Alessandro Vespignani}, {and} {Fillipo
  Menczer}. 2012.
\newblock \showarticletitle{{Competition among memes in a world with limited
  attention.}}
\newblock {\em Scientific reports\/}  {2} (2012), 335.
\newblock
\showISSN{2045-2322}


\bibitem[\protect\citeauthoryear{Weng, Menczer, and Ahn}{Weng
  et~al\mbox{.}}{2013}]%
        {Weng2013}
{Lilian Weng}, {Filippo Menczer}, {and} {Yong-Yeol Ahn}. 2013.
\newblock \showarticletitle{{Virality prediction and community structure in
  social networks.}}
\newblock {\em Scientific reports\/}  {3} (2013), 2522.
\newblock
\showISSN{2045-2322}


\bibitem[\protect\citeauthoryear{Yang and Leskovec}{Yang and Leskovec}{2010}]%
        {Yang2010}
{Jaewon Yang} {and} {Jure Leskovec}. 2010.
\newblock \showarticletitle{{Modeling Information Diffusion in Implicit
  Networks}}. In {\em Proc. ICDM}.
\newblock
\showISBNx{978-1-4244-9131-5}


\bibitem[\protect\citeauthoryear{Yang and Leskovec}{Yang and Leskovec}{2011}]%
        {Yang2011}
{Jaewon Yang} {and} {Jure Leskovec}. 2011.
\newblock \showarticletitle{{Patterns of temporal variation in online media}}.
  In {\em Proc. WSDM}.
\newblock
\showISBNx{9781450304931}


\bibitem[\protect\citeauthoryear{Zhou, Khemmarat, and Gao}{Zhou
  et~al\mbox{.}}{2011}]%
        {Zhou}
{Renjie Zhou}, {Samamon Khemmarat}, {and} {Lixin Gao}. 2011.
\newblock \showarticletitle{{The impact of YouTube Recommendation System on
  Video Views}}. In {\em Proc. IMC}.
\newblock
\showISBNx{9781450304832}


\end{thebibliography}

\received{October 2013}{-}{-}


\end{document}